\documentclass[aps,superscriptaddress,showpacs,nofootinbib,eqsecnum,prd,notitlepage]{revtex4} 

\usepackage{epsf}  
\usepackage{eucal} 
\usepackage{amssymb}
\usepackage{amsmath} 
\usepackage{graphicx} 
\usepackage{bm}
\usepackage{dcolumn} 
\usepackage{epsfig} 
\usepackage{multirow}
\usepackage{hyperref}
\usepackage{subfigure}

\begin{document}

\title{Stability analysis for the background equations for  inflation
  with dissipation and in a viscous radiation bath}

\author{Mar Bastero-Gil} \email{mbg@ugr.es} \affiliation{Departamento
  de F\'{\i}sica Te\'orica y del Cosmos, Universidad de Granada,
  Granada-18071, Spain}

\author{Arjun Berera} \email{ab@ph.ed.ac.uk} \affiliation{SUPA,
  School of Physics and Astronomy, University of Edinburgh, Edinburgh,
  EH9 3JZ, United Kingdom}

\author{Rafael Cerezo} \email{cerezo@ugr.es} \affiliation{Departamento
  de F\'{\i}sica Te\'orica y del Cosmos, Universidad de Granada,
  Granada-18071, Spain}

\author{Rudnei O. Ramos} \email{rudnei@uerj.br}
\affiliation{Departamento de F\'{\i}sica Te\'orica, Universidade do
  Estado do Rio de Janeiro, 20550-013 Rio de Janeiro, RJ, Brazil}

\author{Gustavo S. Vicente} \email{gsvicente@uerj.br}
\affiliation{Departamento de F\'{\i}sica Te\'orica, Universidade do
  Estado do Rio de Janeiro, 20550-013 Rio de Janeiro, RJ, Brazil}

\begin{abstract}

The effects of bulk viscosity are examined for inflationary
dynamics in which dissipation and thermalization are present.
A complete stability analysis is done for the background inflaton
evolution equations, which includes both inflaton dissipation and
radiation bulk viscous effects.    Three representative approaches of
bulk viscous irreversible thermodynamics are analyzed: the Eckart
noncausal theory, the linear and causal theory of Israel-Stewart and a
more recent nonlinear and causal bulk viscous theory. 
It is found
that the causal theories allow for larger bulk viscosities
before encountering an instability in comparison to
the noncausal Eckart theory.  It is also shown that
the causal theories tend to suppress the radiation production  due to
bulk viscous pressure, because of the presence of relaxation effects
implicit  in these theories.  Bulk viscosity coefficients
derived from quantum field theory are applied to
warm inflation model building and an analysis is made of the
effects to the duration of inflation.
The treatment of bulk pressure would also be relevant to the reheating
phase after inflation in cold inflation dynamics and
during the radiation dominated regime, although very little
work in both areas has been done; the methodology developed
in this paper could be extended to apply to these other problems.

\end{abstract}


\maketitle

\section{Introduction}
\label{sec1}

Cosmological inflationary models are usually described by the
evolution of a background scalar field, the inflaton. These models can
be separated in terms of isentropic (cold) \cite{inflation, Linde:1983gd} and
nonisentropic (warm) \cite{bf1,wi} expansion (for earlier related work
see \cite{prewarm}), when regarding the production of 
radiation during the inflation phase. In
cold inflation (CI), the couplings of the inflaton field to other
field degrees of freedom are considered not to influence the inflaton
dynamics in any appreciable way, becoming only relevant at the end of
inflation,  at the time of reheating.  As such, in cold inflation dynamics,
during inflation the universe undergoes a stage of supercooling. In
contrast for warm inflation (WI), the interactions of the inflaton
field with other field degrees of freedom are strong enough to lead to
radiation production at a rate that can  eventually compensate the
supercooling due to expansion.   This will lead to some general
statistical state during inflation.  

The analysis of the quantum dynamics of inflationary models is 
done using the effective action with the
in-in Schwinger closed-time path (or Schwinger-Keldysh) 
method~\cite{calzetta+hu}.
This approach has been used for the study of the quantum dynamics of warm
inflation since its early implementations~\cite{BGR}. 
A review of these quantum field theory methods applied to
warm inflation can be found e.g. in Ref.~\cite{Berera:2008ar}. 
In the context of cold inflation, the Schwinger-Keldysh method
is also commonly used and one primary modification it makes to
the tree level
theory is to add time local corrections to the effective potential due
to radiative effects.  
Another modification that can
be studied by this method is the growth of long wavelength quantum
fluctuations and their effect on the evolution of the
inflaton background and fluctuation modes~\cite{boya1}.
Warm inflation models are those in which
significant time nonlocal contributions also arise from the effective action.
Main consequences of such terms will be dissipation and fluctuation
effects~\cite{calzetta+hu,Gleiser:1993ea}.   
This can very generally lead to a vast variety of particle
production behavior during inflation, thus to various types of
statistical states.  Amongst all of these, the one case which can
be studied most extensively by quantum field theory methods is the
case where the statistical state of the Universe emerging from these
effects is near thermal equilibrium.  Once particle production
occurs during inflation, there are many effects that can follow
and it is a complicated problem.  This is why it is essential to
understand the properties of warm inflation in the near thermal
equilibrium regime, so as to provide insight into the
features that one should also expect when away from this limit.
Of late there have been some efforts to understand
warm inflation models
away from thermal equilibrium \cite{nonthermwi}.  
These
papers have analyzed some of the effects that emerge in
the presence of particle production, but they are at a much
more primary level then the extensive study for the near thermal
equilibrium regime.
In particular, creating a radiation bath of particles
will also then modify the inflaton potential no matter what
statistical state it is in.
Also the nature of the radiation bath of particles will
alter and modify the density perturbations.  In the thermal limit there
are well established methods for examining all these effects,
whereas away from this limit it is a much more difficult problem.

As mentioned above, the case that has been studied
most extensively, since it is the most tractable by quantum field
theory methods, is when the radiation has adequate time to reach close
to a thermal state.  For this case, the radiation production during
inflation manifests itself through an effective dissipation term in
the inflaton background equation. The corresponding dissipative
coefficient can be computed from  first principles in quantum field
theory under appropriate conditions (see
e.g. Refs. 
\cite{BGR,Berera:1998px,Moss:2006gt,Berera:2008ar,BasteroGil:2010pb,BasteroGil:2012cm}).
In particular, the two-stage interaction configuration for quantum
fields proposed in \cite{BR1,BR2}, has been shown can lead to a large
enough dissipative terms, while keeping the quantum corrections to the
inflationary potential  under control and allowing a period of
slow-roll inflation~\cite{br05,Hall:2004zr,MX}. 

In WI, the transfer of energy of the inflaton field to the radiation
bath is mediated by the dissipation term in the inflaton's  evolution
equation. Nevertheless, an additional effect can arise due to inner
couplings  in the radiation fluid itself. Internal decays within the
radiation fluid can make it depart  slightly from thermal
equilibrium. Therefore, the radiation fluid can behave as a non-ideal
fluid and viscosity effects must be taken into
account~\cite{weinberg}.  At the background level, the relevant
viscous effect is due to bulk pressure, since it is 
the only viscous effect
appearing in the background equations for an FRW universe.

The study of bulk viscous effects in cosmology, and in particular in
inflation, has some history to it, focused mainly on the effect of the
bulk pressure as a negative pressure (for a partial sample of the
earlier works on bulk viscous cosmologies, see for example
Ref. \cite{earlierbulk}).  In addition, more recently, there has been
a surge of interest in exploring the effects  of the bulk pressure as
the origin of the present accelerated expansion of the universe (see
e.g. Refs.~\cite{darkenergy}).
Almost all of these past works only use phenomenological forms for
the bulk viscosity.  This paper differs from these past works
since we will apply first principle quantum field theory computed expressions
for the bulk viscosity, based on the calculations found e.g. in Ref.~\cite{jeon},
to warm inflation. There has been very little work
done in studying first principles bulk viscosity expressions
in application to cosmology.  There is one paper we are 
aware of along these lines, which is the first 
reference in \cite{darkenergy};
this paper uses 
quantum field theory derived expressions of bulk viscosity
and shows how they can play the role similar to dark energy.

Our work will examine the consequences of bulk viscosity on warm
inflationary dynamics.  In particular we are interested in determining
the stability conditions of the background equations of warm inflation
when coupled to the bulk viscous radiation bath. Earlier studies
examining the stability properties of the warm inflation equations
include Refs. \cite{HR,MX}, but these studies did not include the
effects of bulk viscous pressure.  Preliminary studies on the
inclusion of bulk viscous pressure have been done in
\cite{pavon},\cite{mimoso}, where only the non-causal theory of Eckart
\cite{Eck} has been used.  Here we extend the stability analysis of
the dynamical warm inflation equations to include bulk viscous
effects.  Moreover our analysis will be done not just for the
non-causal Eckart theory \cite{Eck} but also the causal theories of
bulk pressure \cite{IS2,IS3,Koide},   with a full analysis of the
differences in the resulting dynamics from these different theories.

In the study of viscous effects in cosmology it is common to use
linear expressions to describe the viscous pressure, where it is
assumed that the deviations are close to equilibrium. However, the
viscous pressure can, in principle, take the system far from its
thermodynamical equilibrium, so, we must apply suitable approaches in
order to see if one really needs to use a more robust description,
incorporating nonlinear effects. We consider here three different
theories to describe the viscous pressure: the non-causal theory  due
to Eckart \cite{Eck}, the linear and causal theory of
Israel-Stewart~\cite{IS2,IS3}  and finally we will also use a recent
causal and nonlinear theory proposed by the authors in \cite{Koide},
named by them  \textit{Nonlinear Causal Dissipative Hydrodynamics}
(NLCDH).  There have been other proposals for a nonlinear theory for
the bulk viscous pressure \cite{MaartensMendez,Ottinger},  that make
use of ad-hoc parameters, such as the time where nonlinear effects
become important \cite{MaartensMendez} or functions \cite{Ottinger}
that do not have an immediate interpretation from quantum field
theory, for example being a relaxation time.  In using such approaches
there is no immediate understanding how to associate their parameters
with first principles parameters.  We have considered the theory for
bulk pressure in \cite{Koide} since it utilizes parameters which can
readily be determined from microscopic physics, in particular from
quantum field theory.  The theories for the bulk pressure we analyze
here are more natural to use in field theory model building, where the
dissipation terms, viscosity coefficients and relaxation times are
well defined and can be reliably computed once a specific field theory
model is given.

We study the effects of the inclusion of bulk viscosity in three
commonly used supersymmetric realizations of  warm inflation, the
chaotic, hybrid and hilltop models.  As the bulk viscosity modifies
the background dynamics of warm inflation,  it also changes the
available parameter space for warm inflation, which  is analyzed
here. The bulk viscosity is approximated by the Eckart  description
and we place limits on the couplings of the underlying  particle
physics theory for the validity of this approximation. 

The paper is organized as follows. In Sec. \ref{sec2} we summarize the
relevant expressions for warm inflation and for the theories of bulk
viscous pressure that is relevant for this work. In Sec. \ref{sec3} we
present the respective dynamical systems in each of the three theories
for the bulk viscous pressure we are examining, along with the
resulting stability equations.  In Sec. \ref{sec4}, we present some
numerical results that confirm the stability results.  Also in this
section there is an exposition about the differences in the dynamics
that emerge when considering noncausal and causal theories for the
bulk pressure. In Sec. \ref{sec5} we discuss quantum field
theory derived expressions for the bulk viscosity and conditions
under which they can be applied to cosmology.  
In Sec. \ref{model} we apply the stability analyses to
explore  the parameter space available for warm inflation constructed
from supersymmetric models.  {}Finally,  in Sec. \ref{sec7} we present
our conclusions and final comments.

%%%%%%%%%%%%%%%%%%%%%%%%%%%%%%%%%%%%%%%%%%%%%%%%%%%%

\section{Warm Inflation in a bulk viscous radiation fluid}
\label{sec2}

In the warm inflation scenario, the inflaton evolution is described in
terms of a dissipative equation of motion of the
form~\cite{bf1,wi,BGR,Berera:1998px}

\begin{equation}
\ddot \phi + ( 3 H + \Upsilon ) \dot \phi + V_{,\phi} =0
\,,\label{eominf}
\end{equation}
where $\Upsilon$ is the dissipative coefficient, $V_{,\phi}$ is the
field derivative of the (field and temperature dependent) effective
potential, $H=\dot a/a$ is the Hubble rate of expansion, and $a$ the
scale factor of the Friedmann-Robertson-Walker background metric,
$ds^2 = - dt^2 + a(t)^2 \delta_{ij}dx^i dx^j$.

The dissipation term in Eq. (\ref{eominf}) originates from the
interactions of the inflaton field with other fields. These
interactions effectively render the inflaton as an open system, which
necessarily should give origin to dissipation. The dissipation
coefficient $\Upsilon$ accounts for this effective dissipation of the
inflaton and, given a particle physics  model realization of
inflation, it can be computed microscopically (see e.g. the recent
review~\cite{Berera:2008ar} and references there in).  If enough
dissipation, {\it i.e.}, radiation, can be produced during the
inflationary regime, it can eventually compensate the redshift of the
radiation due to expansion. When this happens, we then term the
inflationary regime as warm inflation.  A particular field theory
model that has been proved successful in generating the WI conditions
is the two-stage mechanism~\cite{BR1,BR2,br05}.  In the two-stage
mechanism, the inflaton field couples to a heavy catalyst field,  and
the latter in turn couples to light degrees of freedom. The evolution
of the background inflaton field excites the catalyst fields, which
then decay into light fields.  Large dissipation has been demonstrated
with this simple mechanism, adequate to drive a warm inflationary
expansion.  When the produced light relativistic particles thermalize,
which is achieved when their creation and scattering times are  less
than the Hubble time in an expanding universe, we can model their
contribution as that of radiation,

\begin{equation}
\rho_r \simeq \frac{\pi^2}{30}g_* T^4 \,,
\end{equation}
where $T$ is the temperature of the thermal bath, and $g_*$ the
effective number of light degrees of freedom\footnote{If not otherwise
  specified, we will take $g_* = 228.75$, the effective no. of degrees
  of freedom for the Minimal Supersymmetric Standard Model, when
  presenting numerical results.}.  

Bulk viscous effects come in to play when the light particles in the
resulting radiation bath in warm inflation can also decay, thus making
the radiation fluid slightly depart from equilibrium. This effectively
renders the radiation as a nonideal fluid and gives origin to a bulk
viscous pressure (there may be other dissipative effects in the
radiation fluid itself, like shear viscous stresses, but these do not
contribute at the background level~\cite{weinberg}, though they may be
relevant in the determination of the spectrum of density
perturbations, as demonstrated recently~\cite{shear}).   In the
presence of a bulk viscous pressure, $\Pi$, the stress-energy tensor
for the radiation fluid is given by~\cite{weinberg,Maartens},

\begin{equation} 
{\cal T}_{\mu\nu}^{(r)}= (\rho_r + p_r + \Pi) u_\mu^{(r)}u_\nu^{(r)} +
(p_r + \Pi) g_{\mu\nu}\,,
\label{Tunu}
\end{equation}
where $\rho_r$ is the radiation energy density, $p_r$ the adiabatic
radiation pressure, $u_\mu^{(r)}$ are the four velocity of the
radiation fluid and $g_{\mu\nu}$ the four-dimensional metric. It
happens then that the bulk pressure enters as a contribution to the
radiation pressure $p_r$, such that we can define in general an
effective pressure for the radiation, ${\tilde p}_r$, given by

\begin{equation}
{\tilde p}_r = p_r + \Pi\,.
\label{tildep}
\end{equation}
The evolution equation for the radiation fluid energy density $\rho_r$
then becomes

\begin{equation}
\dot \rho_r+ 3 H ( \rho_r + \tilde{p}_r) = \Upsilon ( \rho_\phi +
p_\phi) \,, 
\label{eomrad}
\end{equation}
where $p_\phi = \dot \phi^2/2 - V(\phi,T)$, and $\rho_\phi + p_\phi=
\dot \phi^2$.

It is also useful to express this in terms of the entropy density $s$.
{}From the Helmholtz free energy $f= \rho-Ts$, where $f=V(\phi,T)$,
and using $s=-\partial f/\partial T$, the total energy density $\rho$
becomes

\begin{equation}
\rho = \frac{\dot \phi^2}{2} + V(\phi,T) + T s\;,
\label{rhos}
\end{equation}
and the Hubble rate $H$ becomes

\begin{equation}
H^2 = \frac{1}{3 m_{\rm P}^2} \left[  \frac{\dot \phi^2}{2} +
  V(\phi,T) +  T s \right]\;,
\label{Hs}
\end{equation}  
where $m_{\rm P}$ is the reduced Planck mass, 
$m_{P} = 1/\sqrt{8 \pi G} = 2.4\times 10^{18} {\rm GeV}$.
Using also that $p_r = (\gamma-1) \rho_r$ and
that the entropy density $s$ is related to the radiation energy
density by $Ts=\gamma \rho_r$,  Eq.~(\ref{eomrad})  can then be
written in terms of the entropy density as

\begin{equation}
T \dot s+ 3 H ( Ts + \Pi) = \Upsilon \dot \phi^2 \,, 
\label{eoms}
\end{equation} 
where we have used $\gamma = 4/3$, which is valid for a
quasi-equilibrium high temperature thermal bath typical of warm
inflation.  

{}From Eq. (\ref{eoms}), the dynamical effects of the bulk viscosity
can be easily read. Given that the bulk viscous pressure $\Pi$ is
negative, it acts to decrease the radiation pressure, thus enhancing
the effect from the source term on the RHS in the equation for the
entropy density. As a consequence, the entropy density increases, and
therefore the radiation energy density also grows.  On the one hand,
if this bulk pressure term is too large, there is too much radiation
production and the radiation energy density dominates too soon over
the scalar field energy density, thus spoiling inflation. This regime
is called the \textit{unstable regime}. On the other hand, if the bulk
pressure term is controlled to avoid the radiation domination until
the end of inflation, the system is said to be in the \textit{stable
  regime}. In this regime the bulk viscosity gives rise to an
additional negative pressure, and hence, inflation is enhanced.    

To account for the dynamics involving the bulk viscous pressure $\Pi$,
as explained in the introduction, we will consider three different
theories:  the noncausal theory due to Eckart~\cite{Eck}, the linear
and causal theory of  Israel-Stewart (IS)~\cite{IS2,IS3},  and a
recent causal and nonlinear theory, \textit{Nonlinear Causal
  Dissipative Hydrodynamics} (NLCDH), proposed in \cite{Koide}.  The
starting point to build these hydrodynamic theories is the
conservation equations of the stress-energy tensor and the number
density vector $N^\mu=n u^\mu$,

\begin{eqnarray}
 \nabla_\mu {\cal T}^{\mu\nu} = 0\;,\;\;  \nabla_\mu N^\mu = 0\;,
\label{conserv}
\end{eqnarray}
with the additional condition on the 4-entropy,
written in terms of the entropy density $s$, $s^\mu = s u^\mu$,
 that
must satisfy the second law of thermodynamics in its covariant form, 

\noindent
\begin{eqnarray}\label{2ndlaw}
\nabla_\mu s^\mu \geq 0\,.
\end{eqnarray}

The 4-entropy, just like the stress-energy tensor, gains a
contribution coming from the dissipative fluxes,

\begin{eqnarray}\label{eq:ENTROPYNEQ}
s^\mu   = s u^\mu + \frac{Q^\mu}{T}, 
\end{eqnarray}
where $Q^\mu = Q^\mu(N^\mu,{\cal T}^{\mu\nu})$ accounts for the
dissipative fluxes.   
The irreversible thermodynamics comprises of the
dissipative forces to the hydrodynamics variables at equilibrium, the
number density $n$, the energy density $\rho$ and the pressure $p$.
These quantities are able to describe the energy  fluxes in a nonideal
fluid. There are different ways in which this can be done, which lead
to different descriptions for the dissipative fluxes, like  for
example for the bulk pressure. We summarize below the   Eckart, IS and
NLCDH theories for the bulk pressure.

\subsection{Eckart theory for the bulk pressure}

The Eckart theory~\cite{Eck} assumes that the entropy vector $s^\mu$
is linear in the dissipative fluxes. The nonequilibrium contribution
to the entropy vector, $Q^\mu$, to first order, should then be
proportional to the dissipative fluxes. Neglecting dissipative terms
other than the bulk pressure, we then have that

\begin{eqnarray}
Q^\mu \simeq  a(n, \rho) \Pi u^\mu\;,
\end{eqnarray}
where the proportionality factor is obtained from the equilibrium
condition and from the covariant form of the second law of
thermodynamics, Eq. (\ref{2ndlaw}).  This then gives ~\cite{Maartens}

\begin{equation}
T\nabla_\mu s^\mu \simeq  - 3H\Pi \;.
\label{entropyeckart}
\end{equation}
\noindent
To ensure that the second law of thermodynamics, Eq. (\ref{2ndlaw}),
is satisfied and interpreting the term $3H$ in
Eq. (\ref{entropyeckart}) as a dissipative force,  $\chi_{E} = 3H$, 
we impose $\Pi$ to be linear in this force,
expressing bulk viscosity as

\begin{equation}
\Pi = - 3 \zeta_b H\,,
\label{bulkeckart}
\end{equation}
where the proportionality term $\zeta_b\equiv \zeta_b(n,\rho)\geq 0$
is the bulk viscosity coefficient~\cite{weinberg}.

The bulk pressure expressed like Eq. (\ref{bulkeckart}) is a noncausal
theory, {\it i.e.}, the speed of the fluxes propagation is
infinite. The Eckart theory can be considered in some circumstances as
a reasonable approximation for the irreversible thermodynamics. This
may happen, for example,  when sufficiently short relaxation time
scales are involved, otherwise a causal theory would be a much better
choice.  We now turn to the simplest of such a causal theory, the IS
one.

\subsection{Israel-Stewart theory for the bulk pressure}

The IS theory~\cite{IS2,IS3} goes one step further than the Eckart
theory by accounting for second order contributions beyond
equilibrium, by expanding the entropy vector to second order in the
dissipative fluxes. Generically this gives, by again only  considering
the bulk pressure contribution, 

\begin{eqnarray}
s^\mu & \simeq & s u^\mu - \beta_0\Pi^2 \frac{u^\mu}{2T},
\label{EntropySecondOrder}
\end{eqnarray}
\noindent
where $\beta_0 (n,\rho) \geq 0$. {}From the covariant derivative of
the entropy vector,  

\begin{eqnarray}
T\nabla_\mu s^\mu =   -\Pi \left[ 3H  + \beta_0 \dot{\Pi}  +
  \frac{T}{2}\nabla_\mu\left(\frac{\beta_0}{T}u^\mu\right) \Pi\right]
\;,
\label{EntropyIS}
\end{eqnarray}
and from the second law of thermodynamics to be satisfied,
Eq. (\ref{2ndlaw}),  it is imposed again, like in the Eckart case,
that the dissipative fluxes be linear in the dissipative forces. {}For
the bulk pressure $\Pi$ this implies from Eq. (\ref{EntropyIS}) that 

\begin{eqnarray}
\Pi = - \zeta_b \left[ 3H  + \beta_0 \dot{\Pi}  +  \frac{T}{2}
  \nabla_\mu\left(\frac{\beta_0}{T}u^\mu\right) \Pi\right]\;. 
\label{bulkIS1}
\end{eqnarray}
The relation (\ref{bulkIS1}) is analogous to Eq. (\ref{bulkeckart}) in
the  Eckart theory.  The difference here being that, from
Eq.~(\ref{entropyeckart}),  the Israel-Stewart force  is expressed as 

\begin{equation}
\chi_{IS} = 3H  + \beta_0 \dot{\Pi}  +
\frac{T}{2}\nabla_\mu\left(\frac{\beta_0}{T}u^\mu\right) \Pi\;.
\end{equation}
 
By  defining $\tau = \zeta_b \beta_0$, which is interpreted as a
relaxation time for the bulk viscous  processes in the radiation
fluid, then Eq. (\ref{bulkIS1}) can also be rewritten in the form
 
\begin{eqnarray}
\tau \dot{\Pi} + \Pi = - 3\zeta_b H  -  \frac{\zeta_b
  T}{2}\nabla_\mu\left(\frac{\tau}{\zeta_b T}u^\mu\right) \Pi\;, 
\label{bulkIS2}
\end{eqnarray}
and by expanding the derivative in the last term in
Eq. (\ref{bulkIS2}) it can finally be expressed as

\begin{eqnarray}
\tau \dot{\Pi} + \Pi = - 3\zeta_b H  -  \frac{\tau \Pi}{2}\left( 3H +
\frac{\dot{\tau}}{\tau} -\frac{\dot{\zeta}_b}{\zeta_b} -
\frac{\dot{T}}{T} \right)\;,
\label{ISbulk}
\end{eqnarray} 
which is the IS equation for the bulk pressure.

As shown in Ref. \cite{Hiscock:1983zz}, the propagation speed for the
bulk pressure is given by 

\begin{eqnarray}\label{eq:CB}
c_\text{visc}^2 = \frac{\zeta_b}{(\rho + p)\tau}\;,
\end{eqnarray} 
and, thus, for $\tau \neq 0$ there is a finite propagation speed for
the flux,  while for the Eckart theory, where $\tau=0$, it is infinity
(noncausal).  In a quantum field theory description for the radiation
bath, e.g. for example  in the two-stage decay mechanism mentioned
earlier for warm inflation, both the bulk viscosity coefficient
$\zeta_b$ and the relaxation time $\tau$ can be defined unambiguously
and be computed microscopically, just like the dissipation coefficient
$\Upsilon$. In particular, the bulk viscosity coefficient can be
obtained from a Kubo formula for the high-temperature light particles
of the radiation bath~\cite{jeon}, and $\tau$ can be associated with
the respective decay time of these particles, $\tau = 1/\Gamma$, where
$\Gamma$ is the decay width.  {}For the validity of considering a
quasi-equilibrium thermal radiation bath,  we are then required to
impose that

\begin{equation}
\tau H \equiv H/\Gamma < 1\;.
\label{quasieq}
\end{equation} 
Likewise, the assumption of proximity with thermal equilibrium
requires the dissipative fluxes to be small compared to the
equilibrium pressure, 

\begin{equation}
| \Pi | \ll p\;.
\label{Pirho}
\end{equation}

In terms of the quasi-equilibrium conditions (\ref{quasieq}) and
(\ref{Pirho}), the equation for the bulk pressure can also be
expressed in terms of the  Maxwell-Cattaneo equation for $\Pi$
\cite{Landau},

 \noindent
\begin{eqnarray}
\tau\dot{\Pi} + \Pi  = - 3\zeta_b H.
\label{eq:CATTANEO}
\end{eqnarray}
The IS equation for $\Pi$, Eq. (\ref{ISbulk}), can then be seen to
give a correction to the Maxwell-Cattaneo equation, which, after
Eckart, are the simplest equations for the bulk pressure including
relaxation (causal) effects.

\subsection{Nonlinear causal dissipative hydrodynamics theory 
for the bulk pressure}

Next, let us consider the NLCDH theory proposed by the authors of
Ref. \cite{Koide}.  This theory assumes the Eckart force term,
$\chi_{E} = \nabla_\mu u^\mu = 3H$, plus a memory effect, so as to
respect causality. Since this theory in principle makes  no
assumptions about the linearity of the bulk pressure, as it was
assumed in the IS theory for instance, it has been regarded as a
nonlinear theory for the bulk pressure.  The memory effect adds a
relaxation to the system. Recall that the Maxwell-Cattaneo theory
Eq. (\ref{eq:CATTANEO}) is obtained by adding a relaxation time
directly to $\Pi$.  In the NLCDH instead, the memory effect is added
to the quantity $\tilde{\Pi}=\Pi \mathcal{V}$ (where $\mathcal{V}$ is
the volume), which is then integrated in a cell (volume element) of
the fluid.  This is done by imposing the relation $\tilde{\Pi}= -
\zeta_b \mathcal{V} \chi_E$.    Through the addition of the memory
effect, we are lead to~\cite{Koide}

\begin{eqnarray}
\tau\dot{\tilde{\Pi}} + \tilde{\Pi} = -3H\zeta_b \mathcal{V}.
\label{Bulk_V}
\end{eqnarray}
The first term in the above equation results in
$\tau(\dot{\Pi}\mathcal{V} + \Pi \dot{\mathcal{V}})$. After using the
conservation law for the volume in a cell of the fluid, $\nabla_\mu(
u^\mu /\mathcal{V})=0$, it can be shown that~\cite{Koide}

\begin{eqnarray}
\nabla_\mu\frac{ u^\mu}{\mathcal{V}} =
\left(\frac{1}{\mathcal{V}}\right)^{\cdot} +  \frac{1}{\mathcal{V}}
\nabla_\mu u^\mu =
-\left(\frac{1}{\mathcal{V}}\right)^2\dot{\mathcal{V}} +
\frac{1}{\mathcal{V}} \chi =0 \;.
\label{ConservationVolume}
\end{eqnarray}
It follows that $\chi = \dot{\mathcal{V}}/\mathcal{V}$ and then
$\dot{\mathcal{V}}=\chi \mathcal{V} = 3H\mathcal{V}$. From
Eq. (\ref{Bulk_V}),  it then follows the NLCDH equation for the bulk
pressure~\cite{Koide},

\begin{eqnarray}
\tau\dot{\Pi} + \Pi = -3H(\zeta_b + \tau \Pi)\;.
\label{bulkNLCDH}
\end{eqnarray}

\section{The dynamical system of equations for warm inflation in a 
bulk viscous radiation fluid}
\label{sec3}

The relevant equations concerning warm inflation in a bulk viscous
radiation fluid are given by the inflaton evolution equation,
Eq. (\ref{eominf}), the entropy energy density evolution
Eq. (\ref{eoms}), with the bulk pressure $\Pi$  given by: (a) in the
Eckart case Eq. (\ref{bulkeckart}); (b) in the IS case by the
evolution equation (\ref{ISbulk}); and (c) in the NLCDH case by the
evolution equation (\ref{bulkNLCDH}).

Writing the inflaton equation of motion as two first order
differential equations, we have that Eqs. (\ref{eominf}) and
(\ref{eoms}) are equivalently written in the form:

\begin{eqnarray}
\dot{\phi} &=& u, \nonumber\\ \dot{u} &=& - 3Hu - \Upsilon u -
V_{,\phi}  ,\nonumber\\ T\dot{s} &=& - 3HTs - 3H\Pi + \Upsilon u^2
\label{dyn},  
\end{eqnarray}
where

\begin{eqnarray}
H^2 = \frac{1}{3 m_{\rm P}^2}\left( \frac{u^2}{2} + V + Ts\right),\;\;
V=V(\phi,T),\;\; \Upsilon=\Upsilon(\phi,T),
\end{eqnarray}
with expression for the bulk pressure $\Pi$, given by either
Eq. (\ref{bulkeckart}), (\ref{ISbulk}) or (\ref{bulkNLCDH}), depending
on which of the cases is treated.  In all the cases, they depend on
the bulk viscosity coefficient, $\zeta_b \equiv \zeta_b(T)$. In warm
inflation the expected dependence of the dissipation coefficient on
the field and temperature 
is~\cite{Moss:2006gt,Berera:2008ar,BasteroGil:2010pb}
\begin{equation}
\Upsilon = C_\phi \frac{T^c}{\phi^{c-1}} \,,
\label{Upsilon}
\end{equation}
with proportionality factor $C_\phi$ depending on the field content of
the model and the value of the power $c$ depending on the temperature
regime for the different fields involved. {}For example, in the
two-stage decay model for warm inflation, for a radiation bath of
light particles $y$ in the  high temperature regime, $m_y \ll T$, and
for heavy catalyst fields $\chi$ in the low temperature regime,
$m_\chi > T$, we find that  $c=3$
\cite{MX,Berera:2008ar,BasteroGil:2010pb}.  Likewise, the bulk
viscosity coefficient, in this same regime is~\cite{jeon} $\zeta_b
\propto T^3$.  In the following we consider a generic power dependence
$l$ in the temperature for the bulk viscosity coefficient, $\zeta_b
\propto T^l$, similar to that considered for the dissipation
coefficient, Eq. (\ref{Upsilon}).

Treating the variables of the dynamical system in the form of a column
matrix $ \mathbb{X}$, we can express the dynamical system in the
compact matrix form,

\begin{eqnarray}
\dot{\mathbb{X}} = \mathbb{F}(x)\ \mathbb{X}\;,
\label{dynM}
\end{eqnarray}
where for example, for the dynamical system given by Eq. (\ref{dyn}),
$\mathbb{X} = (\phi,u,s)$. In the IS and NLCDH cases we also have the
bulk pressure entering in the system as an additional function,
$\mathbb{X} = (\phi,u,s,\Pi)$.

Writing $x = x_0 + \delta x$, where we assume that $x_0$ is a stable
solution of the system, which here will be taken as the slow-roll
solutions that can be derived directly  from Eq. (\ref{dyn}) (see
below), the equation for the variations in $\delta x$ become

\begin{eqnarray}\label{eq:DELTAX}
\delta \dot{\mathbb{X}} = \mathbb{M}(x_0)\ \delta \mathbb{X} -
\dot{\mathbb{X}}_0\;,
\end{eqnarray}
where

\begin{eqnarray}
\mathbb{M}(x_0) = \frac{\partial \mathbb{F}(x_0)}{\partial x}\;,
\end{eqnarray}
is the Jacobian matrix for the system, evaluated at the $x_0$
solution, and  $\dot{\mathbb{X}}_0$ is a residual force term, which in
general is small and can be neglected~\cite{MX}.   The general
solution of Eq. (\ref{eq:DELTAX}) is of the form 

\begin{eqnarray}
\delta \mathbb{X} = \mathbb{X}_{0} e^{\mathbb{M}(x_0) t}\;,
\label{PertSol}
\end{eqnarray}
and  $\mathbb{M}(x_0)$ must be zero or negative for the system be
stable, {i.e.}, the eigenvalues $\Lambda_i$  of $\mathbb{M}(x_0)$ must
necessarily all satisfy

\begin{eqnarray}
\Lambda_i \leq 0\;.
\label{eigenvalues}
\end{eqnarray}

The stability of the dynamical system can be studied directly in the
time variable,  but it simplifies the analysis, in particular the
determination of the eigenvalues of the Jacobian matrix, if we make a
change of variables and rewrite the dynamical system (\ref{dyn}) using
$\phi$ as the independent variable instead of the time~\cite{MX}.  By
doing this the dimension  of the corresponding system is smaller and
easier to analyze.  In particular, the Jacobian matrix for the
dynamical system (\ref{dyn}) becomes a $2\times2$ matrix; if one
includes the bulk pressure as an additional function to the system, as
in IS and NLCDH cases, it then makes the Jacobian matrix $3\times3$.
The eigenvalues obtained by using $\phi$ rather than time as the
variable makes the analysis much simpler.  As such, using that 

\begin{eqnarray}\label{eq:tu}
\frac{d}{dt} = \frac{d\phi}{dt}\frac{d}{d\phi} = u \frac{d}{d\phi} =
u\ ()',
\end{eqnarray}
where a prime indicates derivative with respect to $\phi$, the
dynamical system Eq. (\ref{dyn}) becomes equivalent to

\begin{eqnarray}
u' &=& - 3H - \Upsilon  - V_{,\phi}u^{-1}\;, \nonumber\\ Ts' &=& -
3HTsu^{-1} - 3H\Pi u^{-1} + \Upsilon u\;,
\label{dynu}  
\end{eqnarray}
together with the corresponding equations for the bulk viscosity,
Eqs. (\ref{bulkeckart}), (\ref{ISbulk}) or (\ref{bulkNLCDH}) (these
last two also transformed to the $\phi$ variable), depending on which
case we are considering.

The general solution of Eq. (\ref{dynu}) is now of the form 

\begin{eqnarray}
\delta \mathbb{X} = \mathbb{X}_{0} e^{\mathbb{M}(x_0) \phi(t)}\;.
\label{PertSol2}
\end{eqnarray}
The stability condition on the eigenvalues $\lambda_i$, which are
the eigenvalues  of  $\mathbb{M}(x_0)$ once $u$ is factorized, depends
now on whether the inflaton field $\phi(t)$  during slow-roll
decreases with time (like in chaotic inflation), so $\lambda_i \geq
0$, or increases with time (like in hilltop inflation), in which case
$\lambda_i \leq 0$. This is an important consideration  when replacing
the time by the inflaton as the independent variable in the dynamical
system. In a $2\times2$ system, stability is ensured once the
determinant is positive and the trace, negative, as is done in
\cite{MX}, \cite{pavon}. However, in a $3\times 3$ system, the case
for the Israel-Stewart and NLCDH descriptions, further information
beyond the trace and determinant is required to guarantee the
stability of the system. That is why in this work we will make use
directly of the eigenvalues for searching for the stability of the
system. Nevertheless, for all the cases we have studied below,
actually only one eigenvalue changes sign at the instability point,
and therefore it would be enough to look at the determinant of the
system. 
 
As mentioned above, the stability analysis is done around the
slow-roll solutions $x_0$.  We study the stability of dynamical system
around the slow-roll solutions,  since they act as formal attractor
solutions for the dynamical  system~\cite{Liddle,HR}.  The slow-roll
solutions are defined by the equations

\begin{eqnarray}
u &=& - \frac{V_{,\phi}}{3H(1 + Q)} \;,\nonumber\\ Ts &=& Qu^2 -
\Pi\;, \nonumber\\ \Pi &=& -3 \zeta_b H \;,
\label{slow-roll}
\end{eqnarray}
where we have omitted the index $0$ that indicates the slow-roll
solutions used in Eq. (\ref{eq:DELTAX}) for simplicity. Note that we
are writing the slow-roll condition for $\Pi$ as the same in all three
cases for the bulk viscous pressure that we have considered in this
work.  This is valid provided that the relaxation time $\tau$ is
sufficiently small, which can  be implicitly assumed by the condition
(\ref{quasieq}). In Eq. (\ref{slow-roll}) we defined $Q \equiv
\Upsilon/(3 H)$, with the Hubble rate $H$, in the slow-roll
approximation as given by

\begin{eqnarray}
H^2 = \frac{1}{3m_{\rm P}^2}V (1 + \kappa) \;,
\label{H2slow}
\end{eqnarray}
where $\kappa = \rho_r/V$. Keeping the radiation energy density in
Eq. (\ref{H2slow})  is justified because in the presence of a bulk
pressure, the radiation energy density does not in general need to be
much smaller than the vacuum energy density in order to have
inflation.  In particular recall that in the presence of a bulk
pressure, the acceleration equation  is

\begin{eqnarray}
\frac{\ddot{a}}{a}  =\frac{1}{6 m_{\rm P}^2}( 2V + 3|\Pi| - 2 \rho_r
)\;,
\end{eqnarray}
which shows that we could in principle have $\rho_r \sim  V$ and
inflation could still be sustained by the bulk
pressure~\cite{MaartensMendez}.
 
The solutions (\ref{slow-roll}) follow if the following conditions on
slow-roll coefficients in warm inflation are satisfied~\cite{MX},

\begin{eqnarray}
\epsilon &=& \frac{m_{\rm P}^2}{2} \left ( \frac{V_{,\phi}} {V}
\right)^2 \ll 1+Q\,, \nonumber \\ \eta &=& m_{\rm P}^2 \left (
\frac{V_{,\phi \phi}} {V}\right)  \ll 1+Q \,, \nonumber \\ \beta &=&
m_{\rm P}^2 \left ( \frac{\Upsilon_{,\phi} V_{,\phi} } {\Upsilon
  V}\right) \ll 1 +Q\,, \nonumber \\ b &=& \frac{T
  V_{,T\phi}}{V_{,\phi}} \ll \frac{Q}{1+Q}  \,,
\label{slowrollcoef}
\end{eqnarray}
where the slow-roll parameter $\beta$  takes into account the
variation of $\Upsilon$ with respect to $\phi$, and  the last
condition for $b$ ensures that  thermal corrections to the inflation
potential are negligible.

Let us now study the stability of the dynamical system of equations in
each of the three cases for the bulk pressure that we are considering.

\subsection{Dynamical system for the Eckart case}

In the Eckart theory the bulk pressure is simply given by
Eq. (\ref{bulkeckart}), $\Pi  = - 3\zeta_b H$. The dynamical system
Eq.~(\ref{dynu}) takes the form,  

\begin{eqnarray}
u' &=& - 3H - \Upsilon  - V_{,\phi}u^{-1} \equiv  f(u, s)\;,
\nonumber\\ s' &=& - 3Hsu^{-1} + 9 \zeta_b H^2 \left(Tu\right)^{-1} +
\Upsilon T^{-1}u \equiv g(u,s)\;.
\label{dyneck}
\end{eqnarray}
The Jacobian matrix $\mathbb{M}$ becomes

\begin{eqnarray}
\mathbb{M}_{\rm Eckart}(x) =
\frac{\partial(f,g)}{\partial(u,s)}\Bigg|_{u = u_0, s=s_0} \equiv
\left(
\begin{array}{ccc}
\partial f / \partial u & \partial f / \partial s\\ \partial g /
\partial u & \partial g / \partial s
\end{array} \right)\Bigg|_{u = u_0, s=s_0} =
\left(
\begin{array}{cc}
\mathcal{A} & \mathcal{B}\\ \mathcal{C} &\mathcal{D}
\end{array} \right)  \;,
\end{eqnarray}
where the matrix elements are evaluated at the slow-roll solutions
Eqs. (\ref{slow-roll}) and (\ref{H2slow}).  The coefficients of the
matrix $\mathbb{M}_{\rm Eckart}$ become

\begin{eqnarray}
\mathcal{A} & = &  \frac{H}{u}\left\{-3(1 + Q) -
\frac{1}{(1+\kappa)^2}\frac{\epsilon}{(1 +
  Q)^2}\right\}\;,\nonumber\\ \mathcal{B} & = &   \frac{H}{s}\Bigg\{ -
3(\gamma -1)cQ + 3(\gamma -1)b(1 + Q)  + \nonumber  \\  &-&
\left.\   \frac{1}{(1+\kappa)^2}\frac{Q\epsilon}{(1 + Q)^2} +
\frac{\sigma}{1+\kappa}\left[ \frac{1}{(1+\kappa)}\frac{Q\epsilon}{(1
    + Q)^2} -  \frac{3}{2}\tilde{\sigma} \right]
\right\}\;,\nonumber\\ \mathcal{C} & = & \frac{Hs}{u^2}\left[6 -
  \frac{1}{(1+\kappa)^2}\frac{\epsilon}{(1 + Q)^2}\right] \left\{1 +
\sigma\frac{\left[6(1 + Q)^2 - 2\epsilon\right]}{\left[6(1 + Q)^2 -
    \epsilon\right]}\right\}\;,\nonumber\\ \mathcal{D} & = &
\frac{H}{u}\left\{ 3(\gamma -1)(c -1) - 3 -
\frac{1}{(1+\kappa)^2}\frac{Q\epsilon}{(1 + Q)^2} + \nonumber
\right.\ \\ &+& \sigma \left.\ \left[ 3(\gamma -1)(c - l) -
  \frac{1}{(1+\kappa)^2} \frac{Q\epsilon}{(1 + Q)^2} +
  \frac{3}{2}\frac{\tilde{\sigma}}{1+\kappa} \right]\right\}\;,
\label{ABCDeck}
\end{eqnarray}
where we have omitted the sub-index ``$0$'' of the slow-roll solutions
and defined the quantities $\sigma$ and  $\tilde{\sigma}$ as

\begin{eqnarray}
\sigma &=& \frac{\Pi}{\gamma
  {\rho_r}}\;,\label{sigma}\\ \tilde{\sigma}  &=&
\frac{\Pi}{V}\;.\label{sigmatilde}
\end{eqnarray}

The expressions simplify considerably in the strong dissipative regime
of warm inflation, $Q \gg 1$ and neglecting the terms proportional to
the slow-roll parameters in Eq. (\ref{ABCDeck}), we obtain

\begin{eqnarray}
\mathcal{A} & = &  -3Q \frac{H}{u}\;,\nonumber\\ \mathcal{B} & = &
3(\gamma -1)(b-c)Q \frac{H}{s}\;,\nonumber\\ \mathcal{C} & = &
\frac{Hs}{u^2}\ 6(1 + \sigma)\;,\nonumber\\ \mathcal{D} & = &
\frac{H}{u}\left\{ 3(\gamma -1)(c-1) - 3  +   \left[ 3(\gamma -1)(c -
  l) + \frac{3}{2}\frac{\tilde{\sigma}}{1+\kappa} \right]\sigma
\right\}\;.
\label{Meck}
\end{eqnarray}

Using (\ref{Meck}), the eigenvalues of $\mathbb{M}_{\rm Eckart}$ are

\begin{eqnarray}
 \lambda_1^{\rm Eckart} &\simeq & -\frac{H}{u}\left[ 3Q  + 6(1+\sigma)
   (\gamma-1)(b-c) \right]  + {\cal O}
 \left(1/Q\right)\;, \label{eige1_eck} \\ \lambda_2^{\rm Eckart}
 &\simeq &  \frac{H}{u}\left\{ 3(\gamma -1)(c-1) - 3  +   \left[
   3(\gamma -1)(c - l) + \frac{3}{2}\frac{\tilde{\sigma}}{1+\kappa}
   \right]\sigma  \right. \nonumber \\ &+& \left. 6(1+\sigma)
 (\gamma-1)(b-c) \frac{}{}\right\} + {\cal O} \left(1/Q\right)
 \;. \label{eige2_eck} 
\end{eqnarray}
Independent of the inflaton dynamics, we then obtain that stability is
assured  when $(u/H) \lambda_i < 0$. 

Note that in the slow-roll regime we have for $\sigma = \Pi/(\gamma
\rho_r)$ that

\begin{equation}
\sigma \simeq \frac{\Pi}{Q u^2 - \Pi} = \frac{\tilde{\sigma}}{\frac{2
    Q}{(1+\kappa)(1+Q)} \frac{\epsilon}{1+Q} - \tilde{\sigma} } \;,
\label{sigmaslowroll}
\end{equation}
where we have used the slow-roll equations for the radiation energy
density and $u$,  Eq. (\ref{slow-roll}).  Note  from the above
equation that in particular we have that $|\sigma| < 1$. Using
(\ref{sigmaslowroll}) in Eqs. (\ref{eige1_eck}) and (\ref{eige2_eck}),
the first eigenvalue will always satisfy the stability condition,
while for the second eigenvalue Eq. (\ref{eige2_eck}),  the stability
condition implies:

\begin{equation}
(c-2b)\left (1 + \sigma \right) + \frac{\gamma}{(\gamma -1)} + l
  \sigma - \frac{1}{2(\gamma - 1)}
  \frac{\sigma\tilde{\sigma}}{1+\kappa} > 0\;,
\label{Eckconditiongamma}
\end{equation}
or, using $\gamma =4/3$, valid for the quasi-equilibrium thermal bath
of warm inflation,

\begin{equation}
(c-2b)\left (1 + \sigma \right) + 4 + l \sigma -
  \frac{3}{2}\frac{\sigma\tilde{\sigma}}{1+\kappa}   >0\;.
\label{Eckcondition}
\end{equation}
The Eq.~(\ref{Eckcondition}) generalizes  the results in ~\cite{pavon},
which were obtained for a constant bulk pressure ($l=0$), by
accounting for temperature dependence\footnote{Our result given by
  Eq. (\ref{Eckcondition}) also corrects Eq. (22) in \cite{pavon},
  where the authors of that reference  mistakenly replaced
  $\tilde{\sigma}$ by $\sigma$.}. Also for $\sigma=0$ and $\tilde
\sigma=0$, the case of zero bulk pressure, we reproduce the results
obtained by Moss and Xiong in \cite{MX}. In ~\cite{MX} the stability
condition was found to be $|c| < 4$. From Eq. (\ref{Eckcondition}), in
the absence of bulk viscosity, we derive instead only the condition
$c> - 4$. We do also obtain the result $c < 4$ if we consider the
eigenvalues in the approximation of very small dissipation $Q \ll 1$,
but this regime is  not the most general situation for warm
inflation~\cite{stochinfl}.

\subsection{Dynamical system for the IS case}

Let us now consider the dynamical system when the bulk pressure $\Pi$
has an evolution  according to the IS theory, Eq. (\ref{ISbulk}).  The
dynamical system, including the evolution equation for the bulk
pressure,  now becomes

\begin{eqnarray}
u' &=& - 3H - \Upsilon  - V_{,\phi}u^{-1} \label{eq:uIS} \equiv  f(u,
s, \Pi)\;,\nonumber\\ s' &=& - 3Hsu^{-1} - 3 H\Pi\left(Tu\right)^{-1}
+ \Upsilon T^{-1}u\label{eq:sIS}  \equiv g(u,s,
\Pi)\;,\nonumber\\ \Pi' &=&  - \frac{\Pi}{\tau}u^{-1} - \frac{3\zeta_b
  H}{\tau}u^{-1} - \frac{\Pi}{2} \Bigg\{ 3Hu^{-1} +  \nonumber \\  &+&
\left[\frac{\tau_{,\phi}}{\tau} - \frac{\zeta_{b,\phi}}{\zeta_b} +
  (\gamma-1)\left(\frac{\tau_{,T}}{\tau} - \frac{\zeta_{b,T}}{\zeta_b}
  - 1\right) \frac{V_{,\phi T}}{s}  \right] + \nonumber \\ &-&
\left.\ (\gamma-1)\left(\frac{\tau_{,T}}{\tau} -
\frac{\zeta_{b,T}}{\zeta_b}  - 1\right)\left[
  3Hu^{-1}\left(1+\frac{\Pi}{Ts}\right) - \frac{\Upsilon u}{Ts}
  \right]   \right\} \equiv  h(u, s,\Pi)\;, \label{dynIS}
\end{eqnarray}
and the Jacobian stability matrix becomes

\begin{eqnarray}
\mathbb{M}_{IS}(x) =
\frac{\partial(f,g,h)}{\partial(u,s,\Pi)}\Bigg|_{u = u_0, s=s_0, \Pi =
  \Pi_0}=  \left(
\begin{array}{ccc}
\mathcal{A} & \mathcal{B} & \mathcal{E}\\ \mathcal{C} & \mathcal{D} &
\mathcal{F}\\ \mathcal{G} & \mathcal{H} & \mathcal{I}
\end{array} \right)\;.
\label{MatrixIS}
\end{eqnarray}

Using the slow-roll solutions, Eqs. (\ref{slow-roll}) and
(\ref{H2slow}), we obtain for the elements of the matrix
$\mathbb{M}_{IS}$ the result

\begin{eqnarray}
\mathcal{A} & = &  \frac{H}{u}\left[-3(1 + Q) -
  \frac{1}{(1+\kappa)^2}\frac{\epsilon}{(1 +
    Q)^2}\right]\;,\nonumber\\ \mathcal{B} & = &  \frac{H}{s}\Bigg\{ -
3(\gamma-1)cQ + 3(\gamma-1)b(1 + Q)  + \nonumber  \\ &-&
\left.\  \frac{1}{(1+\kappa)^2}\frac{Q\epsilon}{(1 + Q)^2} +
\frac{\sigma}{1+\kappa}\left[ \frac{1}{(1+\kappa)}\frac{Q\epsilon}{(1
    + Q)^2} -  \frac{3}{2}\tilde{\sigma} \right]
\right\}\;,\nonumber\\ \mathcal{C} & = & \frac{Hs}{u^2}\left[6 -
  \frac{1}{(1+\kappa)^2}\frac{\epsilon}{(1 + Q)^2}\right] \left(1 +
\sigma\right)\;,\nonumber\\ \mathcal{D} & = &  \frac{H}{u}\left[
  3(\gamma-1)(1+\sigma)c - 3 -  3(\gamma-1) -
  \frac{1}{(1+\kappa)^2}\frac{Q\epsilon}{(1 + Q)^2}
  \right]\;,\nonumber\\ \mathcal{E} & = & 0\;,\nonumber\\ \mathcal{F}
& = & -3 \frac{H}{Tu}\;,\nonumber\\ \mathcal{G} & = &
\frac{HTs}{u^2}\sigma\left\{ \frac{3}{2}+ 3(\gamma-1)\Lambda(1 +
\sigma) +  \nonumber \right.\ \\ &+&
\left.\ \frac{1}{3(1+\kappa)^2}\left[ \frac{1}{\Theta} - \frac{3}{2} -
  \frac{3(\gamma-1)}{2}\Lambda(1 + \sigma)  \right]\frac{\epsilon}{(1
  + Q)^2}  \right\}\;,\nonumber\\ \mathcal{H} & = &
\frac{HT}{u}\sigma\left\{ \frac{(\gamma-1)l}{\Theta} +
\frac{3(\gamma-1)^2}{2}c\Lambda(1+\sigma) -
\frac{3\gamma(\gamma-1)}{2}\Lambda +  \right.\ \nonumber \\ &+&
\left.\ \frac{3(\gamma-1)}{2}b\frac{(1+Q)}{Q}(1+\sigma)\left[
  \frac{Q}{1+Q}\Lambda +  (\gamma-1)\Sigma\right]
+\frac{(\gamma-1)}{2(1+ \kappa)}\frac{\chi}{1+ Q} + \right.\ \nonumber
\\ &+& \left.\ \frac{1}{3(1+\kappa)}\left[ \frac{1}{\Theta} -
  \frac{3}{2} -  3(\gamma-1)\Lambda(1+\sigma) \right]\left[
  \frac{1}{1+\kappa}\frac{Q\epsilon}{(1 + Q)^2} -
  \frac{3}{2}\tilde{\sigma}\right]  \right\}\;,\nonumber\\ \mathcal{I}
& = & \frac{H}{u} \left\{ - \frac{1}{\Theta} - \frac{3}{2} -
\frac{3(\gamma-1)}{2}\Lambda\left[ \sigma + b \frac{(1+ Q)}{Q}(1+
  \sigma)\right] +  \frac{1}{2(1+\kappa)}\frac{\overline{g} -
  \overline{l}}{1+ Q}\right\}\;,
\label{MIS0}
\end{eqnarray}
where we have defined the parameters

\begin{eqnarray}
&&\Theta = \tau H\;,\;\;\; \tilde{l}  = \frac{T \zeta_{b,\phi
      T}}{\zeta_{b,\phi}}\;,\;\;\;  \overline{l}  = \frac{1}{8\pi
    G}\left(\frac{\zeta_{b,\phi} V_{,\phi}}{\zeta_b V}\right)\;,\;\;\;
  \hat{l}=\frac{T \zeta_{b,T T}}{\zeta_{b,T}}\;,\nonumber \\ &&g =
  \frac{T \tau_{,T}}{\tau}\;,\;\;\; \tilde{g}  = \frac{T \tau_{,\phi
      T}}{\tau_{,\phi}}\;,\;\;\; \overline{g}  = \frac{1}{8\pi
    G}\left(\frac{\tau_{,\phi} V_{,\phi}}{\tau V}\right)\;,\;\;\;
  \hat{g}=\frac{T \tau_{,T T}}{\tau_{,T}}\;,\nonumber \\ &&\Lambda =
  1+ l -g\;,\;\;\; \chi = (\tilde{g} -g)\overline{g} - (\tilde{l}
  -l)\overline{l}\;,\nonumber\\ &&\Sigma = (1 + \hat{g} - g)g - (1 +
  \hat{l} - l)l - \Lambda\hat{b}\;,\;\;\; \hat{b} = T V_{,\phi TT}/
  V_{,\phi T}\;. 
\label{parametersIS}
\end{eqnarray}

In the strong dissipation regime  $Q \gg1$ and neglecting terms
proportional to the slow-roll coefficients, Eq. (\ref{MIS0})
simplifies considerably and we obtain

\begin{eqnarray}
\mathcal{A} & = &  -3Q \frac{H}{u}\;,\nonumber\\ \mathcal{B} & = &
3(\gamma -1)(b-c)Q \frac{H}{s}\;,\nonumber\\ \mathcal{C} & = &
\frac{Hs}{u^2}6\left(1 +
\sigma\right),\label{MATRIXISC2}\\ \mathcal{D} & = & \frac{H}{u}\left[
  3(\gamma-1)(1+\sigma)c - 3 -  3(\gamma-1)
  \right]\;,\nonumber\\ \mathcal{E} & = & 0\;,\nonumber\\ \mathcal{F}
& = & -3 \frac{H}{Tu}\;,\nonumber\\ \mathcal{G} & = &
\frac{HTs}{u^2}\sigma\left[ \frac{3}{2}+ 3(\gamma-1) \Lambda(1 +
  \sigma) \right]\;,\nonumber\\ \mathcal{H} & = &
\frac{HT}{u}\sigma\left\{ \frac{(\gamma-1)l}{\Theta} +
\frac{3(\gamma-1)^2}{2}c\Lambda(1+\sigma) -
\frac{3\gamma(\gamma-1)}{2}\Lambda  + \right.\ \nonumber \\ &+&
\left.\ \frac{3(\gamma-1)}{2}b(1+\sigma)\left[ \Lambda +
  (\gamma-1)\Sigma\right]  + \right.\ \nonumber \\ &-&
\left.\ \frac{\tilde{\sigma}}{2(1+\kappa)}\left[ \frac{1}{\Theta}  -
  \frac{3}{2} - 3(\gamma-1)\Lambda(1+\sigma) \right]
\right\}\;,\nonumber\\ \mathcal{I} & = & \frac{H}{u} \left\{ -
\frac{1}{\Theta} - \frac{3}{2} -  \frac{3(\gamma-1)}{2}\Lambda\left[
  \sigma + b(1+ \sigma)\right] \right\}\;.
\label{MIS}
\end{eqnarray}
In terms of Eq. (\ref{MIS}), the eigenvalues of ${\mathbb{M}_{IS}}$
are

\begin{eqnarray}
 \lambda_1^{\rm IS} &\simeq & -\frac{H}{u}\left[ 3Q  + 6(1+\sigma)
   (\gamma-1)(b-c) \right]  + {\cal O}
 \left(1/Q\right)\;, \label{eige1_IS} \\ \lambda_2^{\rm IS} &\simeq &
 \frac{1}{2}\left(\mathcal{D}-\frac{\mathcal{B}\mathcal{C}}
      {\mathcal{A}}+\mathcal{I}\right) + \frac{1}{2}\left[
        \left(\mathcal{D}-\mathcal{I}\right)\left(
        \mathcal{D}-\mathcal{I} -
        2\frac{\mathcal{B}\mathcal{C}}{\mathcal{A}} \right) - 4
        \mathcal{F} \left( \frac{\mathcal{B}\mathcal{G}} {\mathcal{A}}
        - \mathcal{H} \right) +
        \frac{\mathcal{B}^2\mathcal{C}^2}{\mathcal{A}^2} \right]^{1/2}
      + {\cal O}(1/Q)\;,
\label{eige2_IS} \\
\lambda_3^{\rm IS} &\simeq &
\frac{1}{2}\left(\mathcal{D}-\frac{\mathcal{B}\mathcal{C}}
     {\mathcal{A}}+\mathcal{I}\right) - \frac{1}{2}\left[
       \left(\mathcal{D}-\mathcal{I}\right)\left(
       \mathcal{D}-\mathcal{I} -
       2\frac{\mathcal{B}\mathcal{C}}{\mathcal{A}} \right) - 4
       \mathcal{F} \left( \frac{\mathcal{B}\mathcal{G}} {\mathcal{A}}
       - \mathcal{H} \right) +
       \frac{\mathcal{B}^2\mathcal{C}^2}{\mathcal{A}^2}  \right]^{1/2}
     + {\cal O}(1/Q)\;.
\label{eige3_IS}
\end{eqnarray}

For the first eigenvalue above,  $\lambda_1^{\rm IS}$, we see that it
is the same as the Eckart case, $\lambda_1^{\rm Eckart}$, so it
already satisfies the stability requirement that $(u/H) \lambda_i <
0$. Adding all the factors in the two other eigenvalues in
Eqs. (\ref{eige2_IS}) and (\ref{eige3_IS}), using that $\Theta \ll 1$,
we obtain that

\begin{equation}
\lambda_2^{\rm IS} \simeq -\frac{H}{u}\left\{ \frac{1}{\Theta} + 3
|\sigma| \left[ (\gamma-1) l + \frac{|\tilde{\sigma}|}{2(1+\kappa)}
  \right] \right\} + {\cal O}(\Theta)\;,
\label{eige2IS2}
\end{equation}
and it also satisfies the stability requirement. The stability
condition for the IS case then falls on the third eigenvalue
$\lambda_3^{\rm IS}$, Eq. (\ref{eige3_IS}).  This is most easily
expressed by demanding that the product  $\lambda_2^{\rm
  IS}\lambda_3^{\rm IS} > 0$, which then leads to the condition

\begin{eqnarray}\label{eq:CONDSTABIS}
&& \left[  1 + \frac{2\sigma + 3(\gamma-1)\Theta b\Lambda(1 +
      \sigma)^2} {2+ 3\Theta}  \right] c +  \frac{\gamma}{\gamma-1} +
  \frac{2\sigma}{2 + 3\Theta}l  \nonumber \\ && -
  \frac{1}{2(\gamma-1)}\frac{\sigma\tilde{\sigma}}{1+\kappa}\left[
    \frac{2 -  3\Theta}{2 + 3\Theta} -
    \frac{6(\gamma-1)\Theta\Lambda}{2 + 3\Theta}(1 + \sigma) \right] +
  \nonumber \\ &&- \left[ \frac{}{}  4(1+\sigma) + 3\Theta(2+\sigma)
    -3\Theta\Lambda(1+\sigma)^2 + \nonumber \right.\ \\ &&-
    \left.\   3(\gamma-1)\Theta(1+\sigma)(\Lambda+ \sigma  \Sigma) +6
    b (\gamma-1) \Lambda \Theta (1+\sigma)^2 \frac{}{}\right]
  \frac{b}{2+3 \Theta} > 0\;,
\end{eqnarray}
which for  $\gamma =4/3$ becomes

\begin{eqnarray}
&& \left[ 1 + \frac{2\sigma  + \Theta b\Lambda(1 + \sigma)^2}{2+
      3\Theta} \right] c +  4 + \frac{2\sigma}{2 + 3\Theta}l -
  \frac{3}{2} \frac{\sigma\tilde{\sigma}}{1+\kappa}\left[\frac{2 -
      3\Theta}{2 + 3\Theta} -  \frac{2\Theta\Lambda}{2 + 3\Theta}(1 +
    \sigma) \right]  + \nonumber \\ &&- \left[ \frac{}{}  4(1+\sigma)
    + 3\Theta(2+\sigma) - 3\Theta\Lambda(1+\sigma)^2 -
    \Theta(1+\sigma)(\Lambda+\sigma  \Sigma) +2 b \Lambda \Theta
    (1+\sigma)^2 \frac{}{} \right]\frac{b}{2+3 \Theta} >0\;.
\label{IScondition}
\end{eqnarray}
{}From Eq. (\ref{IScondition}), when the relaxation time vanishes,
$\tau=0$, {\it i.e.}, for $\Theta=\tau H \to 0$, we recover the
previous condition Eq. (\ref{Eckcondition}), obtained in the Eckart
theory case.

\subsection{Dynamical system for the NLCDH case}

{}Finally, we will now obtain stability condition for the case of
the NLCDH theory for the bulk pressure. In the NLCDH case, the
evolution equation for the bulk pressure is given by
Eq. (\ref{bulkNLCDH}).  The dynamical system now becomes

\begin{eqnarray}
u' &=& - 3H - \Upsilon  - V_{,\phi}u^{-1} \equiv f(u, s,
\Pi)\;,\nonumber\\ s' &=& - 3Hsu^{-1} - 3 H\Pi\left(Tu\right)^{-1} +
\Upsilon T^{-1}u\equiv   g(u,s, \Pi)\;,\nonumber\\ \Pi' &=&-
\frac{\Pi}{\tau}u^{-1} - \frac{3\zeta_b H}{\tau}u^{-1} -  3H\Pi u^{-1}
\equiv  h(u, s, \Pi)\;. 
\label{dynNLCDH}
\end{eqnarray}
The Jacobian stability matrix is similar to the one in the IS case,
Eq. (\ref{MatrixIS}), but now with the functions $f(u,s,\Pi),\, g(u,s,
\Pi),\,h(u, s, \Pi)$ obtained from the above equation
(\ref{dynNLCDH}).  Using again the slow-roll solutions,
Eqs.~(\ref{slow-roll}) and (\ref{H2slow}), we obtain for the elements
of the matrix $\mathbb{M}_{NLCDH}$ in the NLCDH case,

\begin{eqnarray}
\mathcal{A} & = &  \frac{H}{u}\left\{-3(1 + Q) -
\frac{1}{(1+\kappa)^2}\frac{\epsilon}{(1 +
  Q)^2}\right\}\;,\nonumber\\ \mathcal{B} & = &  \frac{H}{s}\Bigg\{  -
3(\gamma-1)cQ + 3(\gamma-1)b(1 + Q)  +  \nonumber \\ &-&
\left.\ \frac{1}{(1+\kappa)^2}\frac{Q\epsilon}{(1 + Q)^2} +
\frac{\sigma}{1+\kappa}\left[ \frac{1}{(1+\kappa)}\frac{Q\epsilon}{(1
    + Q)^2} -  \frac{3}{2}\tilde{\sigma} \right]
\right\}\;,\nonumber\\ \mathcal{C} & = & \frac{Hs}{u^2}\left[6 -
  \frac{1}{(1+\kappa)^2} \frac{\epsilon}{(1 + Q)^2}\right]\left(1 +
\sigma\right)\;,\nonumber\\ \mathcal{D} & = &  \frac{H}{u}\left\{
3(\gamma-1)(1+\sigma)c - 3 -  3(\gamma-1)-
\frac{1}{(1+\kappa)^2}\frac{Q\epsilon}{(1 + Q)^2}  \right\}
\;,\nonumber\\ \mathcal{E} & = & 0\;,\nonumber\\ \mathcal{F} & = & -3
\frac{H}{Tu}\;,\nonumber\\ \mathcal{G} & = &
\frac{HTs}{u^2}\sigma\left[ 3 + \frac{1}{3(1+\kappa)^2} \left(
  \frac{1}{\Theta} - \frac{3}{2}  \right) \frac{\epsilon}{(1 + Q)^2}
  \right]\;,\nonumber\\ \mathcal{H} & = &
\frac{HT}{u}\sigma\left\{\frac{(\gamma-1)l}{\Theta} +
\frac{1}{3(1+\kappa)}\left( \frac{1}{\Theta}-\frac{3}{2}\right) \left[
  \frac{1}{(1+\kappa)}\frac{Q\epsilon}{(1 + Q)^2}-
  \frac{3}{2}\tilde{\sigma}\right]\right\}\;,\nonumber\\ \mathcal{I} &
= & \frac{H}{u}\left( -\frac{1}{\Theta} -  3  \right)\;,
\label{MNLCDH0}
\end{eqnarray}
with parameters the same as defined in Eq. (\ref{parametersIS}).

Taking again the strong dissipative regime, $Q\gg 1$, and neglecting
terms proportional to the slow-roll coefficients, Eq. (\ref{MNLCDH0})
simplifies to

\begin{eqnarray}
\mathcal{A} & = &  -3Q \frac{H}{u}\;,\nonumber\\ \mathcal{B} & = &
3(\gamma -1)(b-c)Q \frac{H}{s}\;,\nonumber\\ \mathcal{C} & = &
\frac{Hs}{u^2}\ 6\left(1 + \sigma\right)\;,\nonumber\\ \mathcal{D} & =
&  \frac{H}{u}\left[ 3(\gamma-1)(1+\sigma)c - 3 - 3(\gamma-1) \right]
\;,\nonumber\\ \mathcal{E} & = & 0\;,\nonumber\\ \mathcal{F} & = & -3
\frac{H}{Tu}\;,\nonumber\\ \mathcal{G} & = & \frac{HTs}{u^2}\ 3\sigma
\;,\nonumber\\ \mathcal{H} & = &
\frac{HT}{u}\sigma\left[\frac{(\gamma-1)l}{\Theta}  -
  \frac{1}{2}\left( \frac{1}{\Theta}- 3 \right)
  \frac{\tilde{\sigma}}{1+\kappa}\right]\;,\nonumber\\ \mathcal{I}  &
= & \frac{H}{u}\left( -\frac{1}{\Theta} - 3  \right)\;.
\label{MNLCDH}
\end{eqnarray}
The eigenvalues that follow from ${\mathbb{M}_{NLCDH}}$, using
(\ref{MNLCDH}) and considering the high dissipative regime $Q \gg1$,
are still of the form obtained in the IS case,
Eqs. (\ref{eige1_IS}), (\ref{eige2_IS}) and (\ref{eige3_IS}), with
coefficients as given by (\ref{MNLCDH}). Obviously, one of the
eigenvalue is still the same as the one obtained in the Eckart case,
Eq. (\ref{eige1_eck}), while the other two determine the stability
condition for the NLCDH case, similar to Eq. (\ref{eq:CONDSTABIS}),

\begin{eqnarray}\label{eq:CONDSTABNLCDH}
\left( 1 + \frac{\sigma}{1 + 3\Theta} \right) c +
\frac{\gamma}{\gamma-1} + \frac{\sigma}{1 + 3\Theta}l -
\frac{1}{2(\gamma-1)}\frac{\sigma\tilde{\sigma}}{1+\kappa} \frac{1 -
  3\Theta}{1 + 3\Theta}   - \left[ 2 + \sigma +
  \frac{\sigma}{1+3\Theta} \right]b  > 0\;.
\end{eqnarray}
For $\gamma =4/3$ the above equation gives

\begin{eqnarray}
\left( 1 + \frac{\sigma}{1 + 3\Theta} \right) c+ 4 + \frac{\sigma}{1 +
  3\Theta}l -  \frac{3}{2}\frac{\sigma\tilde{\sigma}}{1+\kappa}\frac{1
  - 3\Theta}{1 + 3\Theta}   -  \left( 2 + \sigma +
\frac{\sigma}{1+3\Theta} \right) b > 0\;.
\label{NLCDHcondition}
\end{eqnarray}
If we take the relaxation time as vanishing in
Eq. (\ref{NLCDHcondition}),  $\Theta=\tau H \to 0$, we once again
recover the result Eq. (\ref{Eckcondition}).

\section{Numerical results}
\label{sec4}

In this Section, we study numerically the system of equations for each
of the three cases derived in the previous section. We will verify the
corresponding stability conditions directly through the numerical time
evolution of the corresponding dynamical systems. We will restrict to
models of warm inflation that can be derived from quantum field
theory.  
Moreover to develop the basic ideas, we will
restrict to a dissipation coefficient of the form 
\cite{Moss:2006gt,Berera:2008ar,BasteroGil:2010pb}
$\Upsilon = C_\phi T^3/\phi^2$, which is what is obtained from the
two-stage mechanism in the low temperature regime.
Although our analysis in this paper is restricted to this
particular dissipative coefficient, it can easily be extended
to other dissipative coefficients, with various other types
studied in \cite{BGR,Moss:2006gt,BasteroGil:2010pb,BasteroGil:2012cm}.
The bulk viscosity coefficient will have the
form $\zeta_b = C_b T^3$,  which is obtained for quantum fields  in
the high temperature regime~\cite{jeon} and also is the form
generically considered in hydrodynamics..  This then corresponds to
the case where $c=3$ for the dissipative coefficient in
Eq. (\ref{Upsilon}) and $l=3$ for the bulk viscosity, with $C_\phi$
and $C_b$  (dimensionless) proportionality constants.  Also for
simplicity, we will analyze here the simplest case of  a quadratic
inflaton potential,

\begin{equation}
V= \frac{m_\phi^2}{2}\phi^2\;,
\label{Vquad}
\end{equation}
with a constant relaxation time ($g=\hat{g}=\tilde{g}=\bar{g} =0$ in
the constants defined in (\ref{parametersIS})).  The extension to
other types of potentials, such as a quartic potential or hybrid type
potentials, does not offer additional difficulties and can be easily
implemented.

In the example considered here, the stability conditions
Eqs. (\ref{Eckcondition}), (\ref{IScondition}) and
(\ref{NLCDHcondition}), for the Eckart, IS and NLCDH cases
respectively, reduce to

\begin{eqnarray}
C_{\rm stab}^{\rm Eckart} &=&  3\left(1 + \sigma\right) + 4 + 3\sigma
-  \frac{3}{2}\frac{\sigma\tilde{\sigma}}{1+\kappa}  >
0\;, \label{Ceck} \\ C_{\rm stab}^{IS} &=&  3\left( 1 + \frac{2}{2 +
  3\Theta}\sigma  \right)   +  4 + \frac{6\sigma}{2 + 3\Theta} +
\nonumber \\ &-&
\frac{3}{2}\frac{\sigma\tilde{\sigma}}{1+\kappa}\left[\frac{2 -
    3\Theta}{2 + 3\Theta} -  \frac{8\Theta}{2 + 3\Theta}(1 + \sigma)
  \right]>0\;, \label{CIS}\\   C_{\rm stab}^{NLCDH} &=&  3\left( {1 +
  \frac{\sigma}{1 + 3\Theta}} \right)   + 4 +  \frac{3\sigma}{1 +
  3\Theta} - \frac{3}{2} \frac{\sigma\tilde{\sigma}}{1+\kappa}\frac{1
  - 3\Theta}{1 + 3\Theta}  >0\;.
\label{CNLCDH}
\end{eqnarray}

In all the numerical studies using the inflaton potential
Eq. (\ref{Vquad}), we have kept fixed the values $m_\phi =
\sqrt{8\pi}\times 10^{-6} m_{\rm P}$, the initial value for the
dissipation factor $Q=100$, the initial temperature $T=370\, m_\phi$
and  $\phi(0) = 10.98 \, m_{\rm P}$. The values for $H(0)$ and
$\dot\phi(0)$ follow from the slow-roll conditions. These values
correspond to a proportionality constant $C_\phi \simeq  1.61 \times
10^{8}$ for the dissipation term, which is a typical value found in
the context of WI model building~\cite{BasteroGil:2009ec}.  The value
of the bulk viscosity coefficient $C_b$ is varied and also the value
of the relaxation constant $\tau H = \Theta$, but observing that we
are still  in the region of validity of the thermal radiation bath at
quasi-equilibrium,  $\Theta < 1$.

By letting the system of equations evolve, we determine the critical
value of $C_b$  for which the stability conditions  for each of the
three theories studied here, Eqs. (\ref{Ceck}), (\ref{CIS}) and
(\ref{CNLCDH}), are violated. The corresponding results are given in
Tab. \ref{tab1}.

\begin{table}[h] 
   \centering    \setlength{\arrayrulewidth}{2\arrayrulewidth}
   \setlength{\belowcaptionskip}{10pt} 
   \begin{tabular}{c|c|c}
      \hline $\Theta$ & $\;$ theory $\;$  & $C_b$ \\ \hline & Eckart &
      2232.94 \\ 0.01 & IS      &   2266.48 \\ & NCLDH   &   2300.04
      \\ \hline & Eckart &  2232.94 \\ 0.05 & IS      &   2400.61\\ &
      NCLDH   &  2568.42  \\ \hline & Eckart &  2232.94 \\ 0.1   & IS
      &   2568.23 \\ & NCLDH   &   2903.79 \\ \hline
   \end{tabular}
   \caption{The critical values for the bulk viscosity constant $C_b$.
     The Eckart case is independent of $\Theta$, therefore the value
     for its critical $C_b$ does not change.}
\label{tab1}
\end{table}
We note from the results of Tab. \ref{tab1} that the values for the
bulk viscosity constant $C_b$ for which the stability conditions for
IS and NCLDH cases are violated increases with respect to the Eckart
case as $\Theta$ increases. In Tab. \ref{tab2} we give the
corresponding differences in percentage.

\begin{table}[h]
   \centering   \setlength{\arrayrulewidth}{2\arrayrulewidth}
   \setlength{\belowcaptionskip}{10pt}  
   \begin{tabular}{c|c|c|c}
      \hline theory & $\Theta=0.01$ & $\Theta=0.05$ & $\Theta=0.10$
      \\ \hline IS  & $ 1.5\% $ & $7.5\%$ & $15.0\%$  \\ \hline NLCDH
      & $3.0 \%$ & $15.0\%$ & $30.0\%$ \\ \hline
   \end{tabular}
   \caption{The increase of the critical value of $C_b$ for the causal
     theories  with respect to the Eckart theory.}
\label{tab2}
\end{table}
 
\begin{figure}[h]
\centering \includegraphics[width=7.5cm,height=6.5cm]{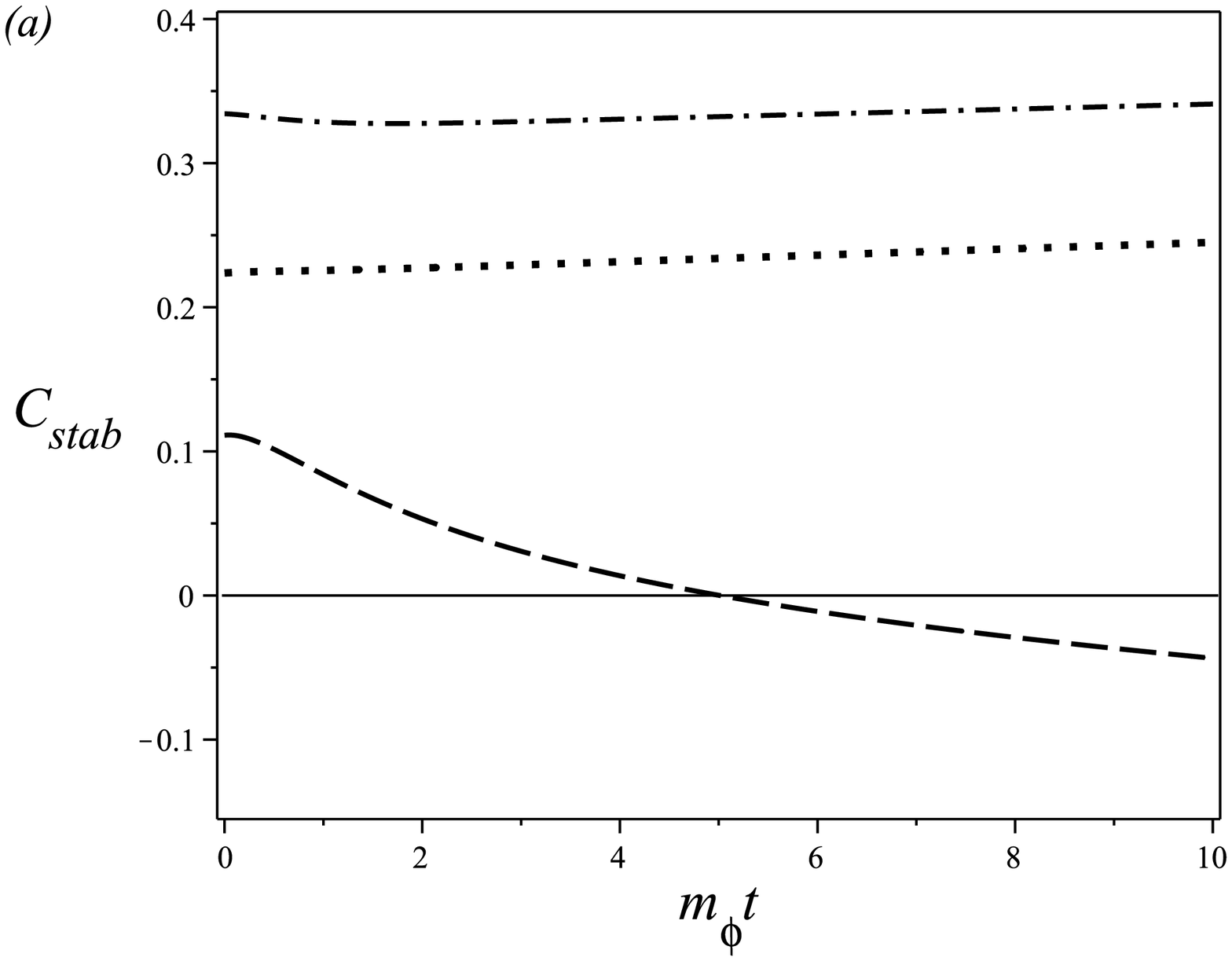}\quad
\includegraphics[width=8cm,height=6.5cm]{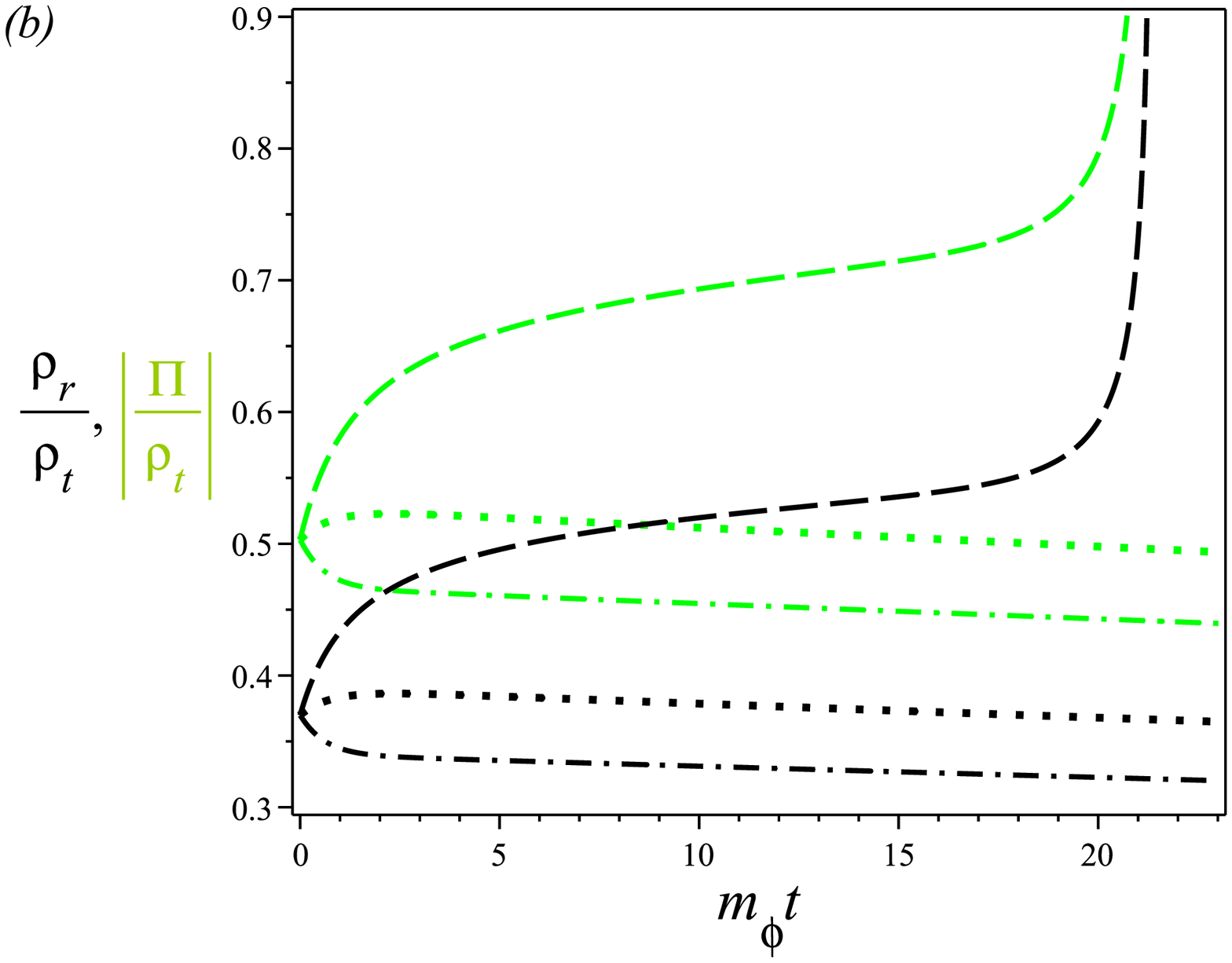}
\caption{The stability condition (a) and the results (b) for the
  radiation energy density, $\rho_r$  (black curves), and bulk
  pressure, $\Pi$ (green curves), normalized by the total energy
  density.  The dashed curves are for the Eckart case, the dotted
  curves are for IS and the dash-dotted  curves are for NLCDH. In all
  cases $\Theta=0.01$ and $C_b=2232.94$.}
\label{fig1}
\end{figure}

\begin{figure}[h]
\centering \includegraphics[width=7.5cm,height=6.5cm]{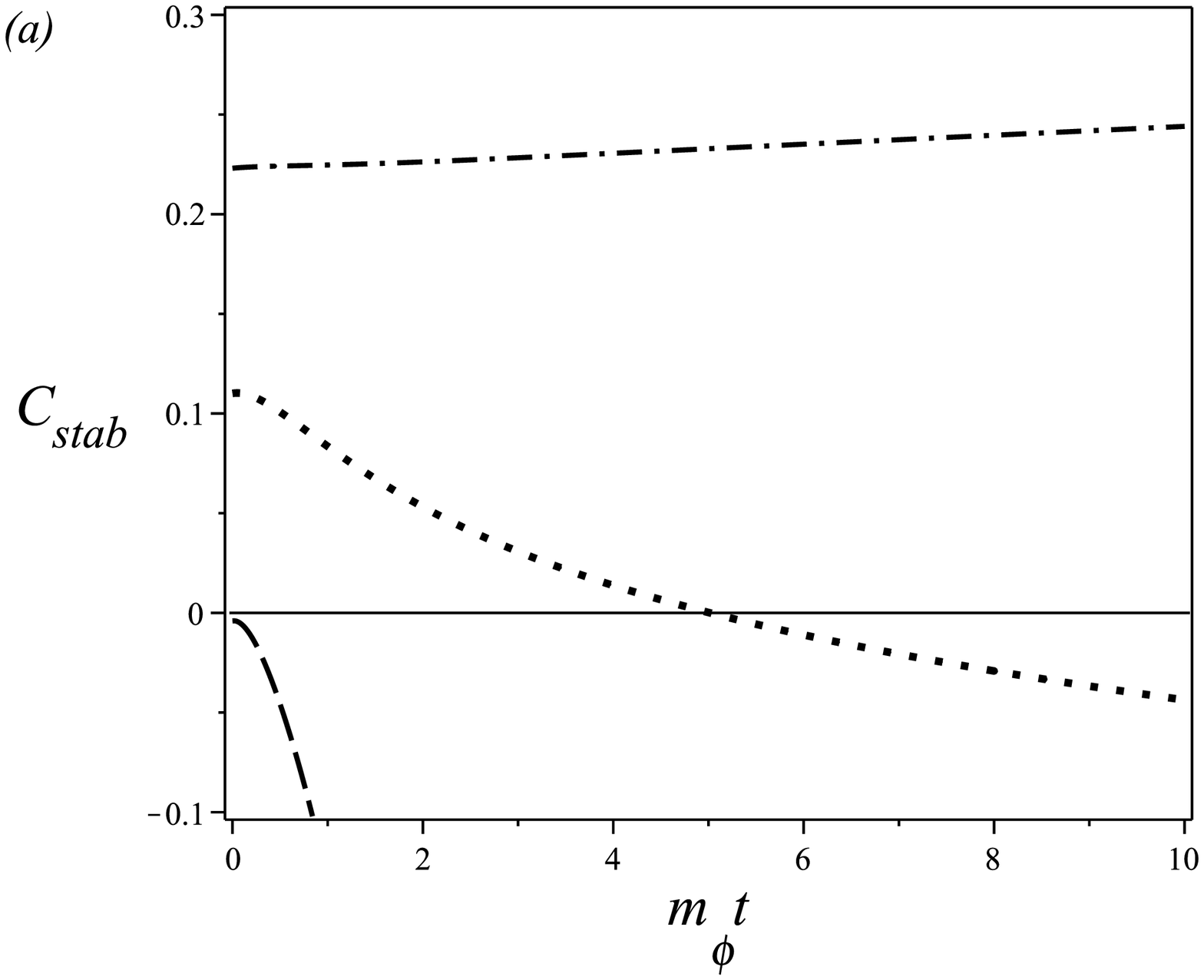}\quad
\includegraphics[width=8cm,height=6.5cm]{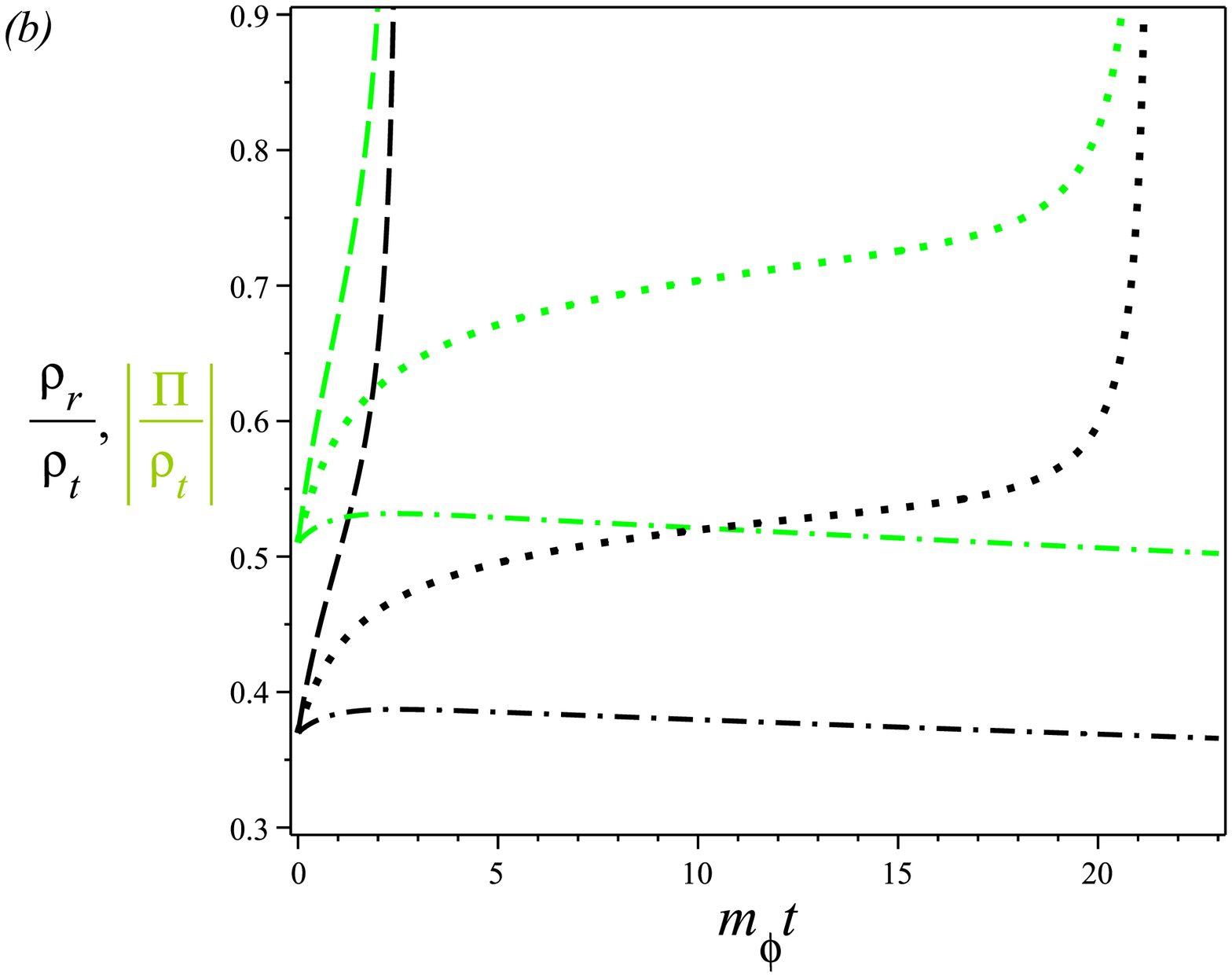}
\caption{The same as in {}Fig. \ref{fig1}, but now for the critical
  value $C_b=2266.48$.}
\label{fig2}
\end{figure}

\begin{figure}[h]
\centering \includegraphics[width=7.5cm,height=6.5cm]{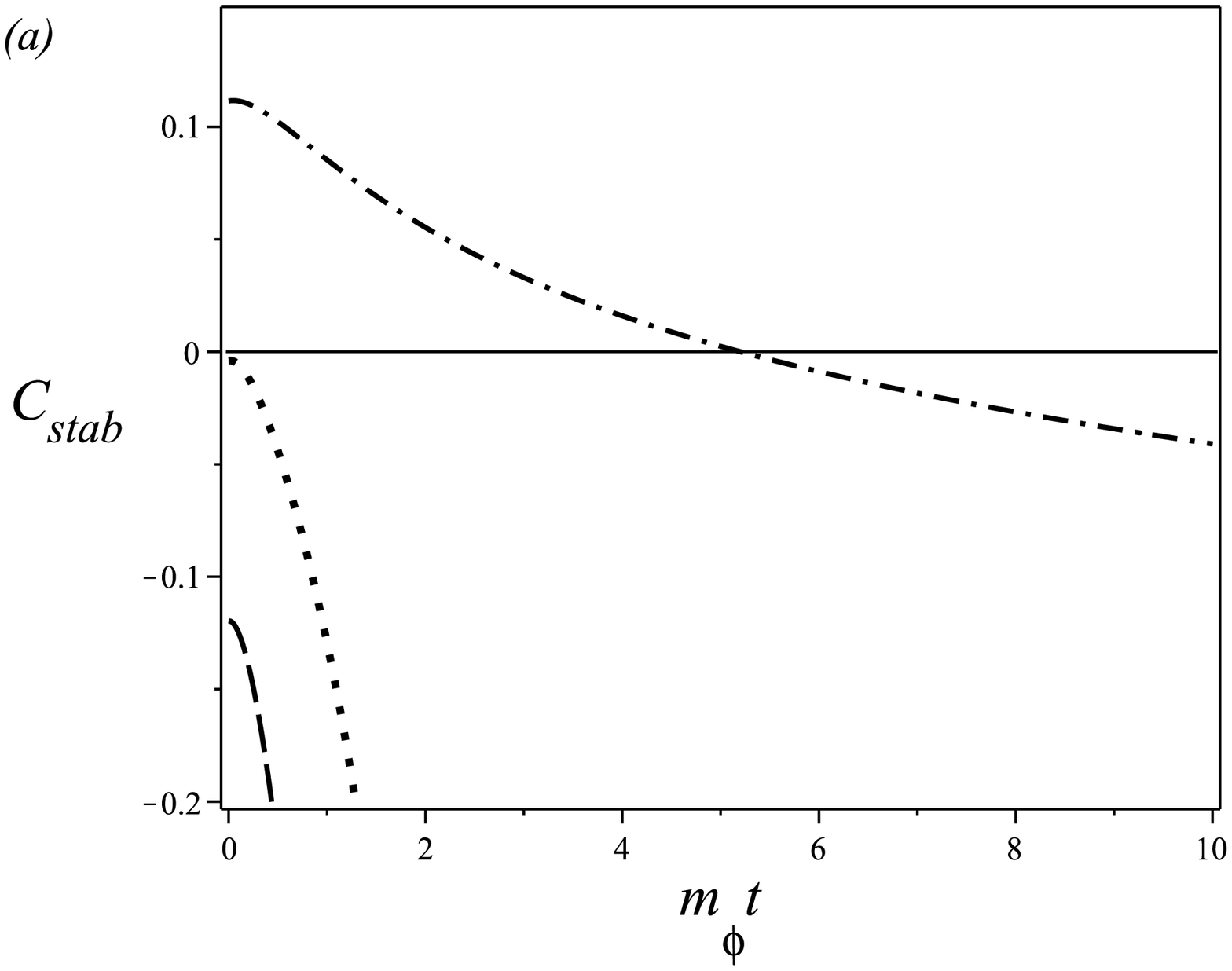}\quad
\includegraphics[width=8cm,height=6.5cm]{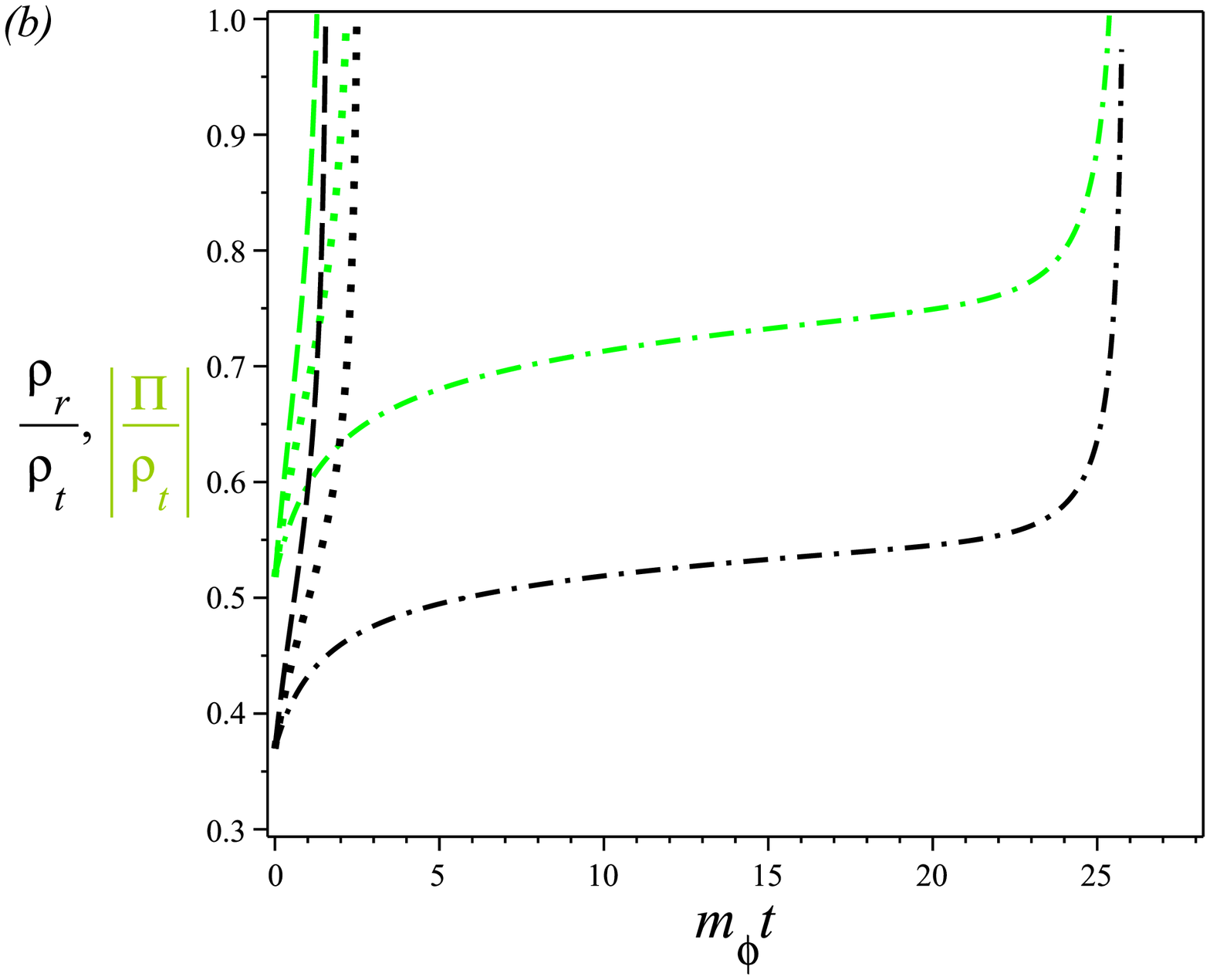}
\caption{The same as in {}Fig. \ref{fig1}, but now for the critical
  value $C_b=2300.04$.}
\label{fig3}
\end{figure}

Once we have the system evolving in time and also allowing the time
dependence for the stability parameters  $C_{\rm stab}^{\rm Eckart}$,
$C_{\rm stab}^{\rm IS}$ and $C_{\rm stab}^{\rm NLCDH}$, by starting
from the initial conditions  given above, we can explicitly check that
the time where Eqs. (\ref{Ceck}), (\ref{CIS}) and (\ref{CNLCDH}), are
violated, is the time where both radiation energy density and the bulk
pressure start to grow exponentially, as expected.  In the
{}Figs. \ref{fig1}, \ref{fig2} and \ref{fig3} we plot the stability
parameters  $C_{\rm stab}^{\rm Eckart}$, $C_{\rm stab}^{\rm IS}$ and
$C_{\rm stab}^{\rm NLCDH}$ along side those for the radiation energy
density and the bulk pressure, as a function of time, for the case of
$\Theta =0.01$ and for the values of critical $C_b$ shown in
Tab. \ref{tab1}.

The results shown in {}Figs. \ref{fig1}, \ref{fig2} and \ref{fig3}
indicate that the time where the stability condition is violated
corresponds to an inflection point in the radiation energy density and
the bulk pressure evolutions. After that time, both   the radiation
energy density and the bulk pressure start to grow and soon after the
dynamics become uncontrollable. The point where the stability
conditions Eqs. (\ref{Ceck}), (\ref{CIS}) and (\ref{CNLCDH}), are
violated, corresponds then to a turnover point in the  evolution of
the dynamical system of equations.

We can also notice from the results for the radiation and bulk
pressure shown in Figs. \ref{fig1}b,  \ref{fig2}b and \ref{fig3}b,
that the causal theories always lead to a  smaller radiation
production when compared to the noncausal case. Among the two causal
theories studied here, the NLCDH theory of Ref. \cite{Koide} gives a
much smaller radiation production from bulk pressure effects than the
IS theory. The differences between the causal theories of IS and NLCDH
are also larger than the noncausal theory of Eckart as the relaxation
time increases, which is clear from the results presented in Tabs.
\ref{tab1} and \ref{tab2}. Even though the difference of the Eckart
theory  for the bulk pressure from that of IS is around the percent
level for a  relatively small relaxation time of $\tau H = 0.01$, it
is twice that  (in percentage) when the NLCDH theory is considered.
This difference between the IS and NLCDH theories can easily be
understood once we compare the Eqs.~(\ref{ISbulk}) and
(\ref{bulkNLCDH}) and  realizes that the relaxation time in the NLCDH
theory appears with an additional factor two on the right-hand-side of
the equation.  Among the theories for the bulk pressure we have
studied here, thus, the NLCDH case is the most robust in terms of
stability.  It allows for relatively larger bulk viscous pressures as
compared to the Eckart and IS cases.

\section{Bulk viscosity coefficients from quantum field theory}
\label{sec5}

The shear and bulk viscosities describe the properties of a system to
return to equilibrium when displaced from it.  As explained in
\cite{jeon}, at the level of particle physics processes, these
viscosities are generally proportional to the mean free path, or
equivalently time, of the relevant scattering process.  The bulk
viscosity is proportional to the mean free path for particle number
changing processes in theories with breakdown of scale invariance.  In
contrast the shear viscosity is proportional to the two-body elastic
scattering mean free path.  Thus the bulk viscosity roughly has the
form
\begin{equation}
\zeta_b \sim \tilde{m}^4 \tau
\end{equation}
where $\tilde{m}$ is a characteristic measure of the violation from
scale invariance in the theory and $\tau$ is the mean free time
between number changing inelastic scattering processes.

Explicit expressions for the bulk viscosity have been calculated for a
self-interacting $\lambda_y y^4$ scalar field theory model in
\cite{jeon}  for different temperature regimes.  The obtained results
for the bulk viscosity relevant for us here are~\cite{jeon},

\begin{eqnarray}
\zeta_b  &\simeq & \left\{
\begin{array}{ll}
& 5.5 \times 10^4 \frac{\tilde{m}_y^4 m_y^2(T)}{\lambda_y^4 T^3}  {\rm
    ln}^2\left[ 1.2465 m_y(T)/T\right] , \;\;  m_y \ll T \ll
  m_y/\lambda_y\\ & 8.9 \times 10^{-5} \lambda_y T^3 {\rm ln}^2
  (0.064736 \lambda_y), \;\; T \gg m_y/\lambda_y ,
\end{array}
\right.
\label{bulksigma}
\end{eqnarray}
where $m_y(T)$ is the scalar $y$ field thermal mass, $m_y^2(T) = m_y^2
+ \lambda_y T^2/24\left[ 1+ {\cal O}(m_y/T) \right]$, and
$\tilde{m}_y^2 \equiv m_y^2(T)-T^2(\partial m_y^2(T)/\partial T^2)
\simeq  m_y^2 - \beta(\lambda_y) T^2/48$, where  $\beta(\lambda_y) = 3
\lambda_y^2/(16 \pi^2)$ is the  renormalization group
$\beta$-function. Note that even for a massless scalar field at
tree-level, $m_y=0$ and thus it is classically scale invariant, the
scale invariance is broken quantum mechanically. In this case, the
$\beta$-function gives a measure of breaking of scale invariance.

The characteristic relaxation time relevant for the bulk viscosity and
that which also enters in the IS and NLCDH formulas for the bulk
pressure,  can be extracted from the result for the bulk viscosity
given by Eq. (\ref{bulksigma}) and the formal expression for it in the
context of the Kubo formula for $m_y \ll T$ and in the relaxation time
approximation~\cite{kapusta},

\begin{equation}
\zeta_b = \frac{1}{T} \int \frac{d^3 p}{(2 \pi)^3}
\frac{\tau(\omega_p)}{\omega_p^2} n(\omega_p) [1+n(\omega_p)]\left[
  \left(\frac{1}{3} - v_s^2 \right) {\bf p}^2 - v_s^2 \tilde{m}_y^2
  \right]^2\;,
\label{kubo}
\end{equation}
where $n(\omega_p) = 1/[\exp(\omega_p/T)-1]$ is the Bose-Einstein
distribution, $\omega_p=\sqrt{{\bf p}^2 + m_y^2(T)}$ and $v_s$ is the
speed of sound for the radiation bath of scalar $y$-particles. Using an
on-shell approximation for the relaxation time, $\tau(\omega_p) \simeq
\tau \, \omega_p/m_y(T)$, where $\tau \equiv \tau(m_y(T)) =$ constant,
and the result for the speed of sound for a self-interacting scalar
field in the high-temperature approximation $m_y \ll T$ \cite{jeon},
$v_s^2 \simeq 1/3 - 5\tilde{m}_y^2/(12 \pi^2 T^2)$, we obtain for the
ratio $\zeta_b/\tau$ the result:

\begin{eqnarray}
\frac{\zeta_b}{\tau} &\simeq & \tilde{m}_y^4 \frac{1}{m_y(T)\, T} \int
\frac{d^3 p}{(2 \pi)^3} \frac{1}{\omega_p} n(\omega_p)
     [1+n(\omega_p)]\left( \frac{5 {\bf p}^2}{12 \pi^2 T^2}
     -\frac{1}{3}\right)^2 \nonumber \\ &\simeq & \frac{\tilde{m}_y^4
       \, T}{18 \pi^2 m_y(T)}  \ln\left(\frac{2T}{m_y(T)} \right)\;.
\label{kubo2}
\end{eqnarray}

Note that for the above results for the bulk viscosity to be applicable
in warm inflation,  it is required that the effective mass for the
scalar $y$ field be larger than the Hubble parameter, $m_y(T) \gg
H$. In this case curvature corrections to the quantum field
expressions defining the bulk viscosity can be neglected and the
Minkowski expression (\ref{bulksigma}) applies. Also, as already
pointed out in the previous sections, a quasi-equilibrium thermal
radiation bath requires that the relevant relaxation time, set by
$\tau$ be also short compared to the Hubble time, $\tau < 1/H$.  These
conditions can be easily meet for warm inflation as verified in the
next section.

\section{Model building}
\label{model}

We have understood how to separate, in general, between the stable and
unstable regimes. Moreover, we have studied the differences between
the non-causal and causal descriptions of the bulk viscosity. In this
Section, we are going to apply this knowledge to some generic
supersymmetric (SUSY) models of warm inflation, namely, the chaotic,
hybrid and hilltop (or new inflation) models.
As an aside, recall that for single field, cold inflation models,
cosmic microwave background radiation (CMB) data~\cite{WMAP} is currently 
able to exclude
many of the chaotic type of models, with potential $V\sim\phi^n$, 
such as with polynomial power $n=3$ (at $95\%$ CL) and
higher powers ($n>3$). Also typical hybrid inflation models
are disfavored by the CMB data for having too large a spectral index
$n_s >1$, while new inflation (hilltop) type cold inflation models seem to be 
more favored by the data~\cite{boya2}. However, these conclusions do
not apply to warm inflation models \cite{BasteroGil:2009ec}, and in
particular to large field inflation models with dissipation
\cite{pavon,stochinfl}, for which the predictions related to the
primordial spectrum (spectral index, running of the spectral index,
non-gaussianity parameter) are different from those of standard cold
inflation. In this paper we have only focused on the background
evolution and so these observational details are
not directly relevant.  A detailed study of the primordial spectrum,
that includes dissipation and viscous effects and the observational
consequences, will be presented elsewhere. 
The goal of this Section is to study the
parameter space of these basic inflation models 
in the presence of bulk viscosity.

As discussed in previous sections, the bulk viscous pressure decreases
the radiation pressure and so allows the source term creating
radiation to be more effective.  Thus assuming thermalization, it
raises the temperature. Given that the dissipative coefficient depends
on the temperature, the bulk viscosity enhances it; and therefore, the
inflaton can slow-roll down its potential with lower values of the
dissipative factor $C_{\phi}$. Hence, we expect an enlargement of the
parameter space in regions of low $C_{\phi}$, where warm inflation is
not allowed in the absence of bulk viscosity. In addition, in the
regions of the parameter space where warm inflation is allowed without
bulk viscosity, its effect is to produce more e-folds of inflation.

One of the main issues that need to be faced in warm inflation are the
quantum and thermal corrections to the inflaton potential, that tend
to spoil the required flatness of the potential needed for
inflation. We invoke SUSY to avoid the quantum corrections, which at
the same time does not cancel the time non-local terms that give rise
to the dissipative terms. The pattern of interactions leading to warm
inflation is given by the superpotential:

\begin{equation}
 W = W(\Phi) + g\Phi X^2 + hXY^2\;,
\label{superW}
\end{equation}
where $\Phi$, $X$ and $Y$ denote superfields. The regime most
studied for warm inflation is where the $X$ field, being
coupled to the inflaton field, is massive during inflation, and
therefore thermal corrections to the inflaton mass are Boltzmann
suppressed as long as  $m_{\chi}>T$, where $\chi$ is the scalar
component of the superfield $X$ and $m^2_{\chi}=2g^2\phi^2$.  Even
though the $X$ field in (\ref{superW}) is at low temperatures,  we can
still have the $Y$ field at high temperatures, e.g., $m_y \ll T$,
where $y$ is the scalar component of $Y$ and $m_y$ its mass.  In this
case, thermal dissipation of the inflaton field mediated by $X$
decaying into $Y$ is not Boltzmann suppressed \cite{BasteroGil:2010pb}
and can lead to a warm inflation regime.  We denote this regime as the
low-$T$ regime and it leads to a dissipative coefficient
of the form  
\cite{Moss:2006gt,Berera:2008ar,BasteroGil:2010pb}
$\Upsilon = C_\phi T^3/\phi^2$.
Other regimes and dissipative coefficients could also be studied,
but we will develop the basic methodology in this
paper with this example.

Warm inflation imposes another condition on the
temperature, $T>H$. This condition roughly separates warm inflation
from cold inflation, since for $T<H$ dissipation effects are negligible.
We also must take into account the thermodynamical condition
$|\Pi/p_r|<1$, as the hydrodynamic descriptions of the bulk viscosity
that we are using treat the viscous pressure as a perturbation to the
equilibrium one.

Therefore, we define the available parameter space for warm inflation as the
regions where the following conditions hold:

\begin{enumerate}
 \item $\epsilon_{H}=-\dot{H}/H^2 < 1,$ which is the standard
   condition for the accelerated expansion,
 \item $\rho_{\phi} > \rho_{r},$ which prevents the radiation energy
   density to dominate,
 \item $T/H>1$, which leads to non-negligible dissipation effects,
 \item $\phi/T \gtrsim 10$, which is the low-T condition for
   $g=\mathcal{O}(1)$,
   \item $|\Pi/p_{r}|< 1$, which is the condition for the hydrodynamic
     description to hold.
\end{enumerate}
These conditions need to hold for at least 40 e-folds to solve the
flatness and horizon problems. In standard cold inflation, this value
is taken to be 50-60 e-folds; nevertheless, the exact number depends
on the scale of inflation and the initial post-inflation temperature.
As warm inflation usually decreases both quantities, it is very
reasonable to demand 40 e-folds.  Condition 2 is controlled by the
stability conditions found in the previous section. {}From
system of equations (\ref{slow-roll}), we can relate the rest of the conditions
with slow-roll parameters (\ref{slowrollcoef}). Condition 1 during
slow-roll simply is:

\begin{equation}
\epsilon_{H}=\frac{\epsilon}{1+Q} < 1\,.
\end{equation} 
The evolution of the ratio $T/H$, in the slow-roll regime, with
respect to the number of e-folds is  given by:

\begin{equation}
 \frac{d\ln (T/H)}{dN_e}=\frac{2(1+\sigma)}{1+Q+6Q(1+\sigma)}\left (
 \frac{2+4Q}{1+Q}\epsilon - \eta +
 \frac{1-Q}{1+Q}\frac{m_P}{\phi}\sqrt{2\epsilon} \right).
\end{equation}
 $\Pi/p_r$ is directly related to $T/H$, 
 \begin{equation}
	 \left|\frac{\Pi}{p_{r}}\right| = \frac{270C_b}{\pi^2
           g_{*}}\frac{H}{T}.
	 \label{piTH}
 \end{equation}
The evolution of $\phi/T$ by:

\begin{eqnarray}
 \frac{d\ln(\phi/T)}{dN_{e}} &=&
 \frac{-1}{\left[1+Q+6Q(1+\sigma)\right]} \left[ \frac{3 +
     4 \sigma  + (1 +
     2\sigma )Q}{1+Q}\epsilon  
%\right. \nonumber
%   \\ &-& \left. 
-2(1+\sigma)\eta +
   \frac{3 + 2\sigma+ 
     (5+ 4 \sigma ) Q }{1+Q }\frac{m_P}{\phi}\sqrt{2\epsilon}
   \right ].
\end{eqnarray}
In addition to these equations, the slow-roll evolution of the field
$\phi$ is given by:

\begin{equation}
	\frac{d\phi/m_P}{dN_{e}}=-\frac{\sqrt{2\epsilon}}{1+Q}.
\end{equation}
For completeness, we also show the evolution of $Q$:

\begin{equation}
 \frac{dQ}{dN_{e}}=\frac{Q}{1+Q+6Q(1+\sigma)}\left[10\left(1 +
   \frac{6}{5}\sigma\right)\epsilon - 6(1+\sigma)\eta +
   8\left(1+\frac{3}{4}\sigma\right)\frac{m_P}{\phi}\sqrt{2\epsilon}\right]\;.
\end{equation}
These results generalize the ones obtained in \cite{BasteroGil:2009ec}
for the case with no bulk viscosity. Note the difference in the
notation between our $\sigma$ and the $\sigma_\phi$ defined there,
which have replaced here  by
$\sqrt{2\epsilon}(\phi/m_P)$. Nevertheless, as we have shown in
$(\ref{sigmaslowroll})$, $|\sigma|\lesssim 1$ and then the results in
\cite{BasteroGil:2009ec} concerning whether the conditions increase or
decrease during the evolution are still valid.

The last question before entering in the particular details of each
model is how we are treating the bulk viscosity. In this Section we
will use the non-causal description of the bulk viscosity, i.e. the
Eckart theory, and place limits on the validity of this
approximation. As discussed previously, the Eckart description is a
good approximation for low values of $\Theta=\tau H$. The Hubble
parameter is given by (\ref{Hs}) and the relaxation time $\tau$ is
obtained from  Eq. (\ref{kubo2}). Using for example the first
expression  for the bulk viscosity in  Eq. (\ref{bulksigma}) (the
second expression for the bulk viscosity in Eq. (\ref{bulksigma}) can
be easily seen as obtained from the first, when neglecting the
temperature independent terms in $m_y(T)$ and in $\tilde{m}_y$), we
obtain that

\begin{equation}
\tau \approx 9.77 \times 10^6 \frac{m_y^3(T)}{\lambda_y^4 T^4},
\end{equation}
where for the superpotential $(\ref{superW})$, we have that
$\lambda_y=6h^2$.  Based
on the previous discussions, we consider the Eckart approximation to
be valid when $\Theta\lesssim 10^{-2}$, which translates into  (using
that in the high-temperature limit $m_y(T) \approx hT/2$)

\begin{equation}
 h\gtrsim 10  \left(\frac{H}{T}\right)^{1/5}\;.
\label{hcond}
\end{equation}
One of the conditions for warm inflation is that $T/H\gtrsim 1$,
therefore, we can easily arrange the condition (\ref{hcond}) to be
satisfied when deep in the warm inflationary regime, particularly in
the strong dissipative regime, which can also allow for perturbative
values for the coupling $h$. The effect of including a causal
description of the viscosity  is to produce a lower value of the bulk
viscous pressure than in the non-causal case with the same coefficient
$C_{b}$. The consequence is a shift around the $5\%$ level in the
entire parameter spaces shown in the next subsections to higher values
of $C_{b}$. This is explicitly verified below for the specific inflaton
potential models we have studied.

%%%%%%%%%%%%%%%%%%%%%%%%%%%%%%%%%%%%%%%%%%%%%%%%%%%%%%%%%%%%

\subsection{Chaotic model}

We first consider a chaotic inflation potential of the form:

\begin{equation}
 V(\phi)=\frac{\lambda}{4}\phi^4 \,
\label{lambdaphi4}
\end{equation}
where we have used $\lambda=10^{-14}$. However this parameter is only 
relevant for the amplitude of the power spectrum, which we are not 
interested in here. Using again that $\Upsilon= C_\phi T^3/\phi^2$,
for the potential (\ref{lambdaphi4})
the slow-roll parameters, Eq. (\ref{slowrollcoef}), are given by

\begin{equation}
 \eta=12\left(\frac{m_{P}}{\phi}\right)^2 \,,\;\;\;
 \epsilon=\frac{2}{3}\eta \,,\;\;\; \beta=-\frac{2}{3}\eta\;.
\end{equation}
Therefore, the value of the field decreases during inflation,
meanwhile, the dissipative ratio $Q$ and $T/H$ both increase. Hence, once
the condition on $T>H$ is fulfilled initially, it is always
satisfied. The ratio $\phi/T$ decreases, but we have checked that it
always remains above $10$ as long as the other conditions are
fulfilled. The parameter $\epsilon_{H}$ increases during inflation
and, as a consequence, warm inflation ends when
$\epsilon_{H}=1$. {}Finally, the condition $\rho_\phi > \rho_r $ is
controlled by the stability condition (\ref{Eckcondition}). It is
only necessary to check that the stability condition is positive at
the beginning of inflation, as it does not change sign during the
evolution. This last statement is true for all the models studied.

\begin{figure}[h]
\centering \includegraphics[scale=0.16, angle=-90]{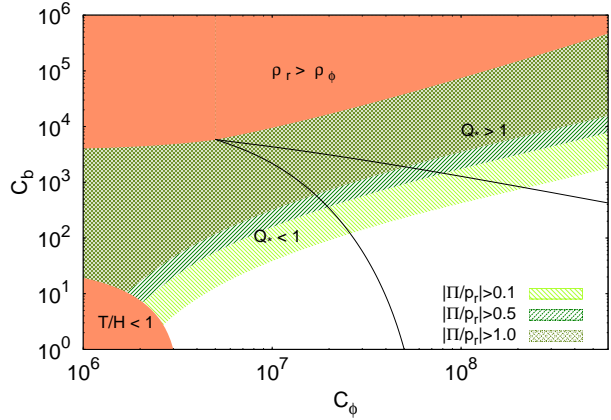}
\caption{Parameter space for the chaotic model. The green regions are
  excluded because of the violation of the condition written in the
  plot. The lines separate regions where $Q_*<1$ and $Q_*>1$
  respectively. In the region between them, we can have both $Q_*<1$
  and $Q_*>1$ for different values of $\phi(0)$.}
\label{figch}
\end{figure}

The available parameter space is shown in {}Fig.~\ref{figch}. For
completeness we have included the parameter space excluded for
different values of the hydrodynamic condition $|\Pi/p_{r}|$, namely
$0.1$, $0.5$ and $1$. We observe that the enlargement of the parameter
space in regions of low $C_\phi$ is not very efficient. In particular,
the minimum value of $C_{\phi}$ is reduced from $2.1\times 10^6$ up to
$1.6\times 10^6$. These values have been confirmed by using the NLCDH
description of the bulk viscosity, Eq. (\ref{bulkNLCDH}), with a
constant $\tau$ fixed by imposing initially the values $\Theta=0.01$,
$0.9$.  The NLCDH description reduces the bulk viscous pressure
associated to a $C_b$ value when $\Theta$ grows, which  means that
higher values of $C_b$ are allowed before the condition $|\Pi/p_r|$ is
violated. However, at the same time, for the same value of $C_\phi$,
higher values of $C_b$ are  required to avoid the $T/H <1$ exclusion
region, therefore the effects cancel each other and the minimum value
of $C_\phi$ is independent of the initial value $\Theta$. 

We have also found that the condition $|\Pi/p_r|$ is the most restrictive one
in almost the full parameter space and that the instability regime
studied in the previous sections is far beyond the limit imposed by
this condition. In the analysis we have fixed the initial values such
that the exclusion regions are the least stringent, that is, the upper
region is as high as possible and the bottom one, as low as
possible. We have fixed initial conditions in this way for the three
models studied.

\begin{figure}[h]
\centering \includegraphics[scale=0.16, angle=-90]{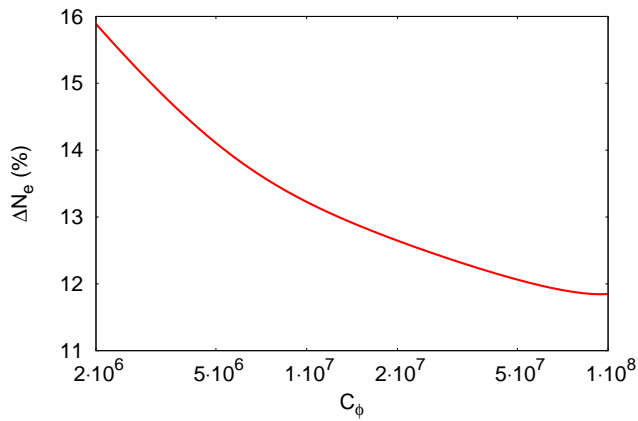}
\caption{Maximum enhancement in the number of e-folds for different
  values of $C_\phi$ for the model with quartic potential. The maximum
  value is obtained for $C_b$=8.25.}
\label{nech}
\end{figure}

In addition, we separate with black lines the regions where the
dissipative ratio at horizon crossing $Q_*$ is always greater than one
from the regions where it is always less than one. In the region
between them, $Q_{*}$ can be either greater or less than one,
depending on the initial value of $\phi$. In the regions that were not
allowed in the absence of bulk viscosity, its main role is to produce
enough e-folds of inflation.  In regions allowed with no bulk
viscosity, the total number of e-folds is increased. We quantify this
effect in terms of the percentage difference $\Delta N_e$, defined as 

\begin{equation}
	\Delta N_e = \frac{N_{e}^{\text{bulk}} -
          N_{e}^{\text{no-bulk}}}{N_{e}^{\text{no-bulk}}}\times 100,
\end{equation}
where $N_{e}^{\text{bulk}}$ is the maximum number of e-folds obtained
with bulk viscosity for a certain $C_{\phi}$  and
$N_{e}^{\text{no-bulk}}$ is the equivalent without bulk viscosity. The
results are  shown in {}Fig. \ref{nech}.

The bulk viscosity can enhance the number of e-folds through two
mechanisms. The first one is, for a given initial value of the field,
to reduce the redshift of the radiation energy density by decreasing
its total pressure. This effect produces an increase in the value of
$Q$, which goes as $\rho_r^{3/4}$ and is related to the number of
e-folds via

\begin{equation}
	N_{e}=\int_{\phi_{end}}^{\phi(0)}{\frac{3H^2(1+Q)}{V_{,\phi}}d\phi},
	\label{efolds}
\end{equation}
where $\phi(0)$ is the initial value of the field and $\phi_{end}$,
the value of the field at the end of inflation. Hence, the increase in
$Q$ leads to a larger number of e-folds. However, we have checked that
this mechanism is subdominant in the quartic potential, providing an
efficiency up to $2\%$. 

The second mechanism allows to increase the initial value of the
field.  {}From Eq. (\ref{efolds}) it can be seen that this produces
more e-folds by increasing the integration interval.  In the absence of
bulk viscosity, there   is an upper limit on the value of the field
because of the condition $T/H > 1$. As the bulk viscosity increases the
value  of $T$, it pushes upwards this limit and, hence, it is possible
to use larger values of the field. However, there is a bound to this
effect, provided by the condition $|\Pi/p_r| < 1$, which translates into a
new upper limit to $\phi(0)$.  The decrease in this effect shown in
{}Fig.~\ref{nech} is because the limits on the initial value of the
field grow at different rates with $C_{\phi}$, with the former growing
faster, which causes the difference on the limits to decrease with
$C_\phi$ and this, in turn, reduces the difference between the viscous
and non-viscous cases.  {}From Eq. (\ref{piTH}) we can obtain the value
of $C_{b}$ that maximizes this mechanism. Using the values
$|\Pi/p_r|=1$, $T/H=1$ and $g_*=225.78$ as an example, we find that
$C_{b}=8.25$. This argument is model independent, so that we find the
same value of $C_b$ in the three models studied and independently of
the value of $C_\phi$.

\subsection{Hybrid models}

We consider now small field models of inflation with an inflationary
potential given by:

\begin{equation}
 V=V_{0}\left[1+\frac{\delta}{2}\left(\frac{\phi}{m_{P}}\right)^2
   \right],
\label{hybrid}
\end{equation}
where $V_0$ is the scale and $\delta$ a model parameter. Here we have
used $V_0=10^{-8}m_P^4$ and study the dynamics for two representative
values for the parameter $\delta$.  The slow-roll parameters, in the
case of the inflaton potential given by Eq.~(\ref{hybrid}), are now
given by

\begin{equation}
 \eta = \delta \,,\;\;\;
 \epsilon=\frac{\delta^2}{2}\left(\frac{\phi}{m_{P}}\right) ^2
 \,,\;\;\; \beta = -2\eta.
\label{slhy}
\end{equation}

During the evolution the value of $\phi$ decreases, while the value of
$Q$ increases. The evolution of the value of $T/H$ depends on the
value of Q: for $Q<1$, $T/H$ increases and for $Q>1$, it decreases;
$\phi/T$ always decreases, and $\epsilon_{H}$ is not relevant in this
model because it is suppressed by a factor $(\phi/m_{P})^2$, which is
usually very small due to the fact that $\phi \ll m_{p}$. Hence,
$\epsilon_{H}$ is always below one. As a result, inflation ends
because the conditions imposed either on $T/H$ or $\phi/T$ are  violated, 
or because
the field reaches its critical value.

\begin{figure}[h]
\centering \subfigure[$\delta=0.1$]  { \centering
  \includegraphics[scale=0.16, angle=-90]{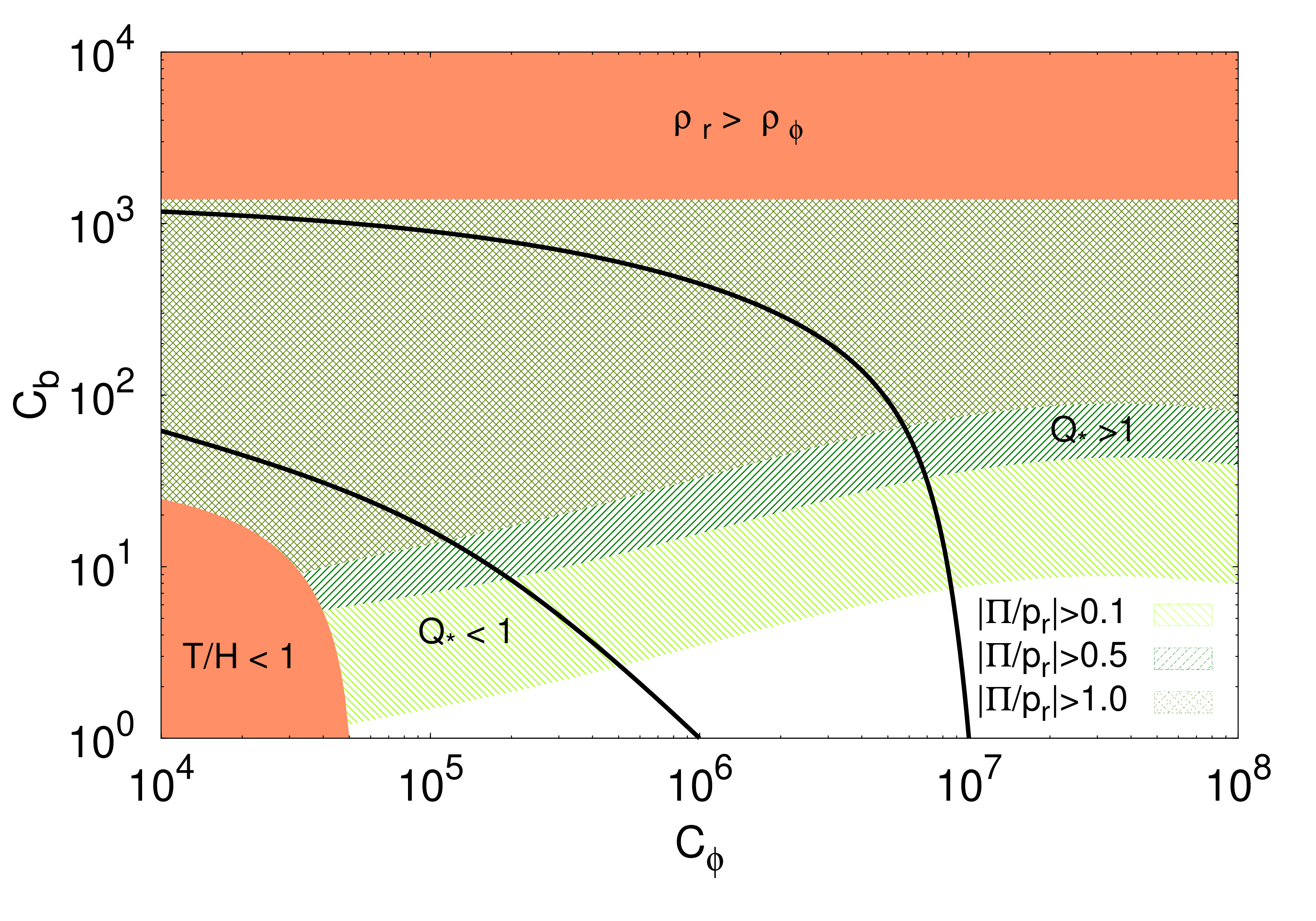}
\label{fighy01}
}\subfigure[$\delta=10$] { \centering \includegraphics[scale=0.16,
    angle=-90]{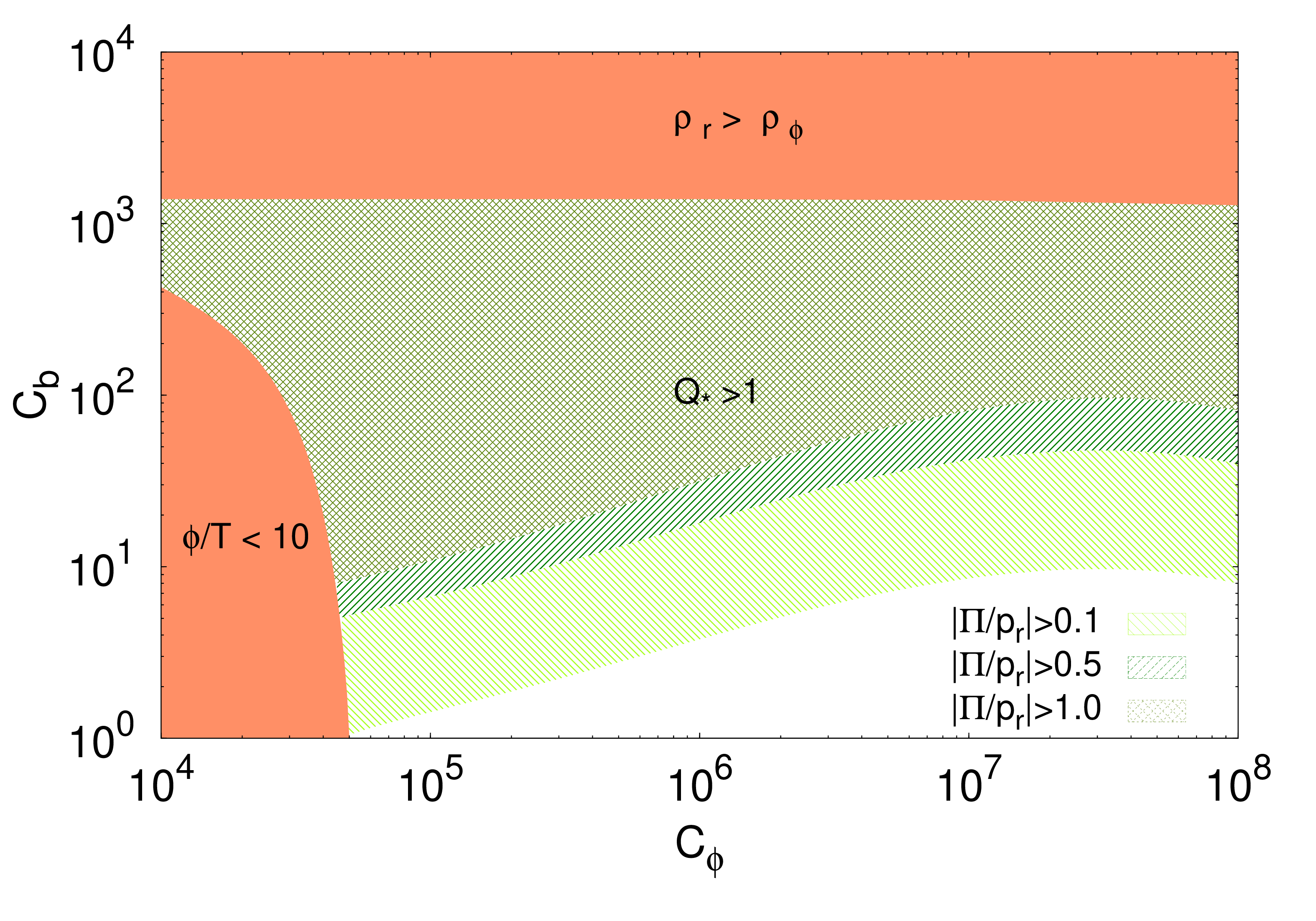}
\label{fighy10}
}
\caption{Parameter space for the hybrid models with $\delta=0.1,10$
  The green regions are excluded because of the violation of the
  condition written in the plot. In the left plot the lines separate
  regions where $Q_*<1$ and $Q_*>1$ respectively. In the region
  between them, we can have both $Q_*<1$ and $Q_*>1$ for different
  values of $\phi(0)$. In the right plot, $Q_{*}$ is always above 1.}
\label{fighy}
\end{figure}

In {}Fig.~\ref{fighy}, the parameter space of the hybrid models with
$\delta=0.1$ (left panel) and $\delta=10$ (right panel) are
plotted. In the left panel, as in the previous model, the black lines
separate regions with different value of the dissipative ratio at
horizon crossing. However, in the right panel, the dissipative ratio
is always above one. {}For a higher value of $\delta$ we can
maintain $\eta/(1+Q)$ below one only for $Q_*>1$. In the left panel
the bottom excluded region is forbidden for the same reason as in the
chaotic model. Nevertheless, in the right panel we find that the
excluded region is forbidden by the $\phi/T > 10$ condition. This is caused
again by the higher value of $\eta$. The 
parameter $\delta$ in Eq. (\ref{hybrid}) measures the curvature of the
potential. Thus, for higher values of $\delta$ the field evolves
faster. As a result, the condition on $\phi/T$ is reached first than in the
case for smaller values for the parameter $\delta$.

\begin{figure}[h]
\centering \includegraphics[scale=0.16, angle=-90]{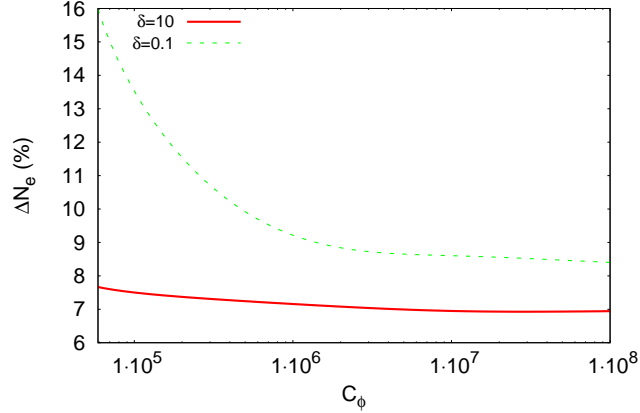}
\caption{Maximum enhancement in the number of e-folds for different
  values of $C_\phi$ in the case of the inflaton potential  given by
  Eq.~(\ref{hybrid}). The maximum value is obtained for $C_b=8.25$.} 
\label{nehy}
\end{figure}

The minimum value of $C_{\phi}$ in this case of inflation with the
hybrid type of potential, Eq.~(\ref{hybrid}), is reduced from
$3.5\times10^4$ up to  $2.6\times10^4$, for $\delta=0.1$, while for
$\delta=10$ it is reduced from $5\times10^4$ up to
$4.1\times10^4$. Making use of the $NLCDH$ description of the bulk
with a constant $\tau$ fixed by imposing the initial values
$\Theta=0.001$, $0.9$, we found that the minimum value $C_\phi$ is
independent of the initial choice of $\Theta$. The effect on the
number of e-folds is shown in {}Fig.~\ref{nehy}. As in the previous
case, the constant field mechanism is subdominant,  with an efficiency
of around a $3\%$ for the $\delta=10$ case and a $6\%$ efficiency for
the $\delta=0.1$ case.

\subsection{Hilltop models}

We now consider new inflation hilltop type of models, which are
characterized by an inflaton potential given by

\begin{equation}
 V=
 V_{0}\left[1-\frac{|\delta|}{2}\left(\frac{\phi}{m_P}\right)^2\right] +
 \cdots,
 \label{hilltop}
\end{equation}
where the dots account for higher-order terms and
$V_0=10^{-8}m_P^4$. This is a potential similar to the two previous
ones, but with a negative squared inflaton mass. The slow-roll
parameters are still given by those in Eq.~(\ref{slhy}), but with the
change $\delta \to -\delta$.  In these models $\phi$, $\phi/T$, $T/H$
and $\epsilon_{H}$ increase during the evolution, while $Q$
decreases. Inflation ends when the field reaches a large enough value,
so that higher-order terms in the potential start contributing and
$\epsilon_H$ becomes greater than one.

\begin{figure}[h]
\centering \subfigure[$\delta=0.1$] { \centering
  \includegraphics[scale=0.16, angle=-90]{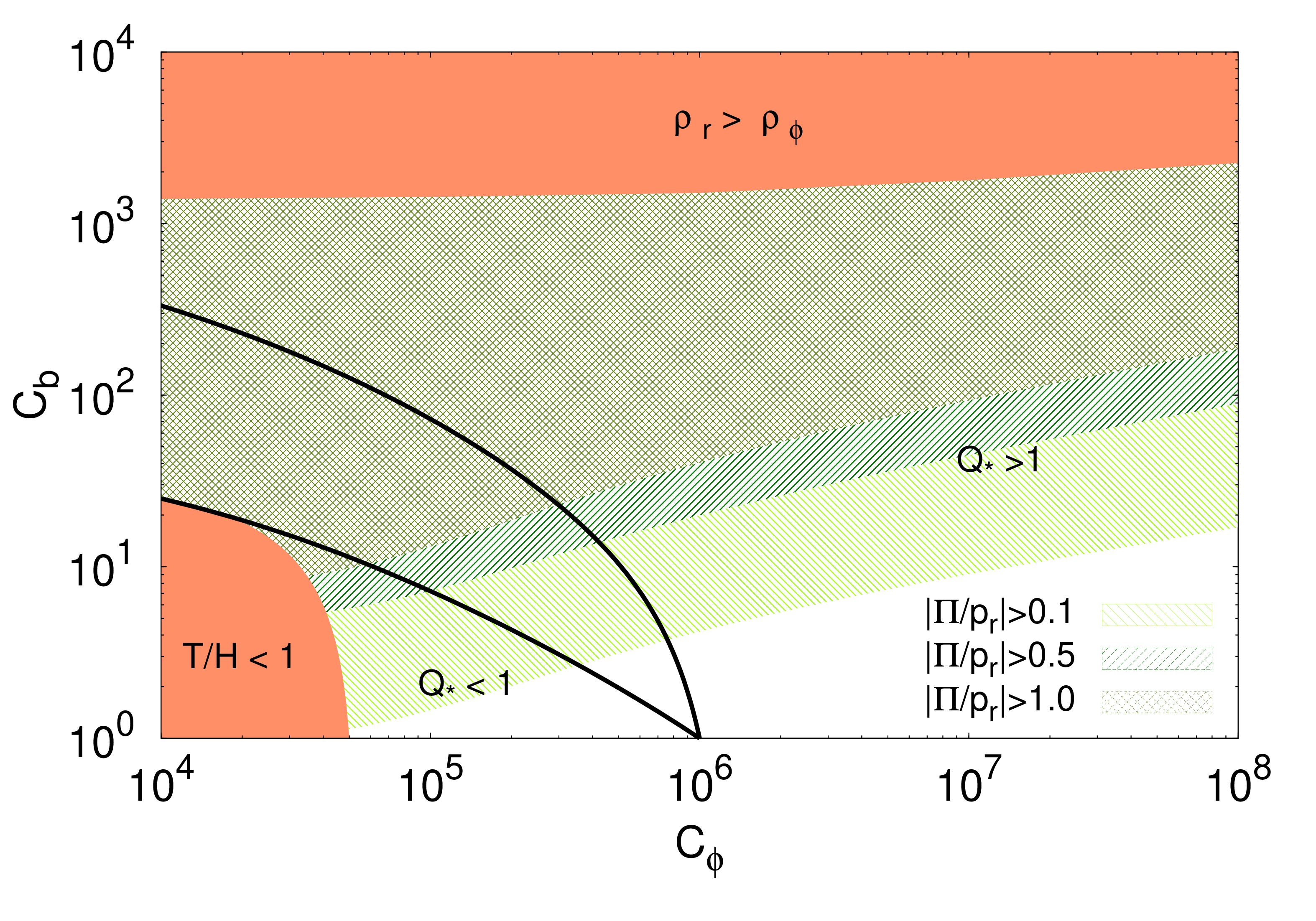} 
}\subfigure[$\delta=1$] { \centering \includegraphics[scale=0.16,
    angle=-90]{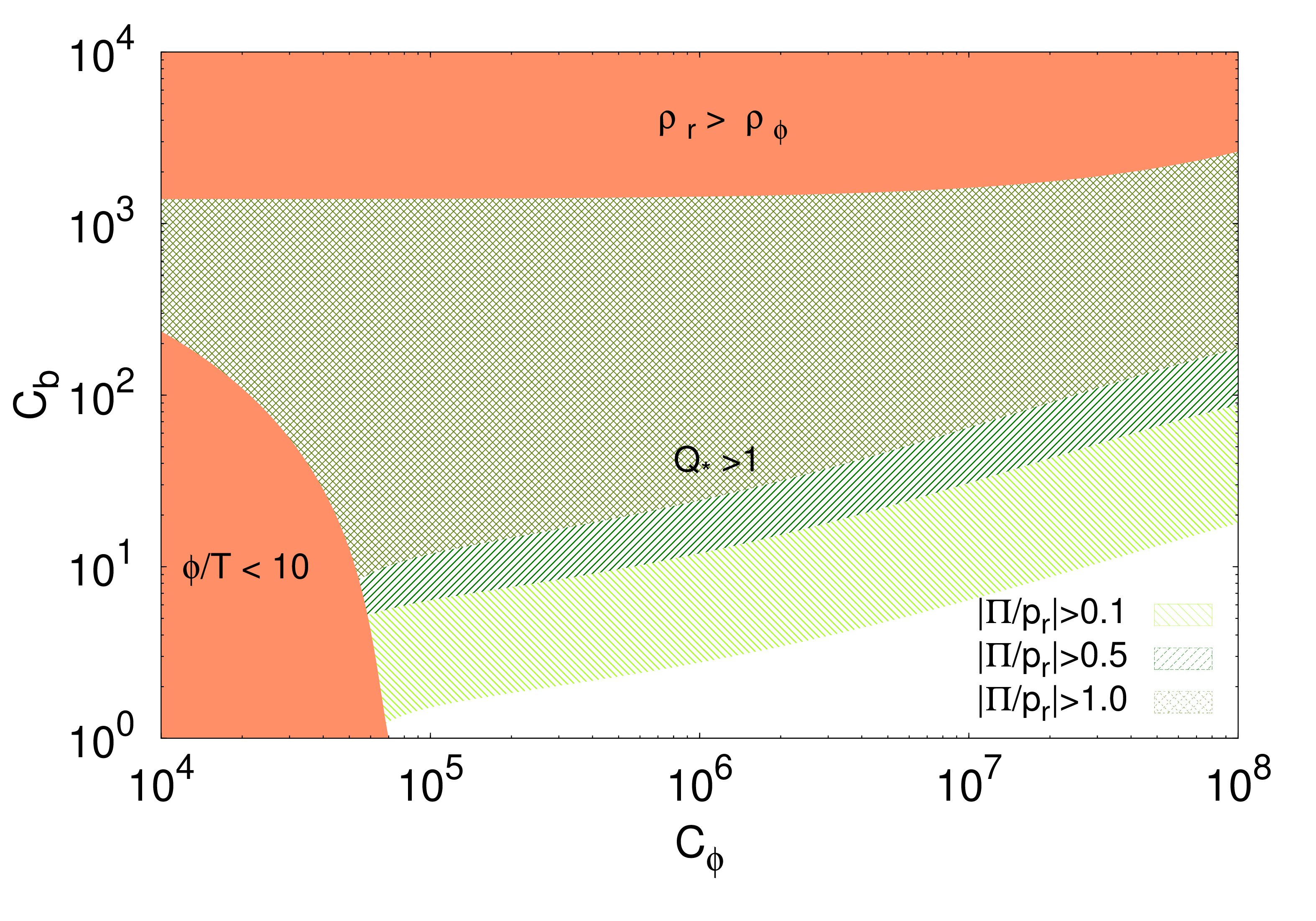}
\label{fighi1}
}
\caption{Parameter space for the hilltop models with $\delta=0.1,1$.
  The green regions are excluded because of the violation of the
  condition written in the plot. In the left plot the lines separate
  regions where $Q_*<1$ and $Q_*>1$ respectively. In the region
  between them, we can have both $Q_*<1$ and $Q_*>1$ for different
  values of $\phi(0)$. In the right plot, $Q_{*}$ is always above 1.}
\label{fighi}
\end{figure}

The parameter space for the hilltop model Eq. (\ref{hilltop})  is
shown in  {}Fig.~\ref{fighi}, for the cases of $\delta=0.1$ (left
panel) and $\delta=1$ (right panel).  Once again, in the left panel
the black curves separate regions with $Q_*$ greater or less than one
at horizon crossing, and in the right panel, the dissipative ratio at
horizon crossing is always greater than one due to the large value of
$\eta$. 

\begin{figure}[h]
\centering \includegraphics[scale=0.16, angle=-90]{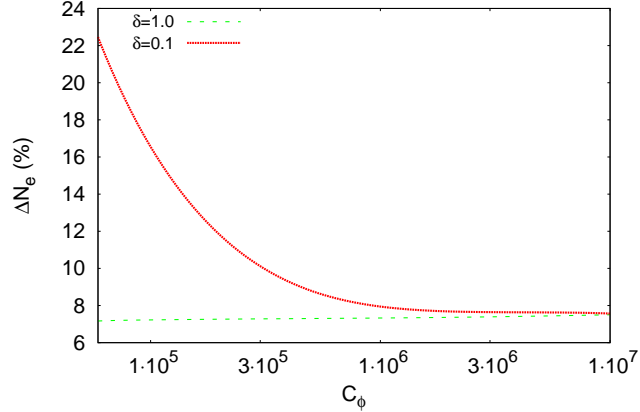}
\caption{Maximum enhancement in the number of e-folds for different
  values of $C_\phi$, for the case of the inflaton  potential
  Eq.~(\ref{hilltop}). The maximum value is obtained for $C_b=8.25$.} 
\label{nehi}
\end{figure}

Now, the minimum value of $C_{\phi}$ is reduced from $3.3\times10^4$
up to $2.5\times10^4$ for $\delta=0.1$ and from $5.9\times10^4$ up to
$5.1\times10^4$ for $\delta=1$.  We have checked  these values with
the NLCDH description of the bulk with a constant $\tau$ fixed by
imposing the initial values $\Theta=0.001$, $0.9$. The effect on the
number of e-folds is  shown in {}Fig.~\ref{nehi}. As in the other
cases studied, the constant field mechanism is subdominant with
efficiencies around a $5\%$ and a $3\%$ for the $\delta=1$ and the
$\delta=0.1$ cases respectively.  The initial value mechanism works
reversely compared with the previous potentials. Here it allows to use
lower initial values of the field, however, as the value of the field
grows in this case, this reduction implies  an increase of the
integration interval in Eq. (\ref{efolds}). In addition, note that in
this potential, the lower the value of the field, the larger the value
of $H$ and thus, the lower is the ratio $T/H$. 
This argument also applies  to the
$|\Pi/p_r|$ condition, therefore, there are lower limits to $\phi(0)$
rather than upper ones.
 
\subsection{The bulk viscosity coefficient}

Let us now briefly discuss the expected values for the bulk
viscosity coefficient for a warm inflation model described
by Eq. (\ref{superW}). In the context of the model described by 
Eq. (\ref{superW}), for the
relevant temperature regime for warm inflation, characterized by
$m_\chi > T$ and $m_y \ll T$,  the dominant contribution to the bulk
viscosity  comes from the radiation thermal bath composed of the
high-temperature particles, {\it i.e.} the $y$ particles and given by
Eq. (\ref{bulksigma}).  We can see from the expression for the bulk
viscosity Eq. (\ref{bulksigma}) that the larger values for the bulk
coefficient $C_b = \zeta_b/T^3$ appears in the intermediate
temperature regime, $m_y \ll T \ll m_y/\lambda_y$, where the $y$
particles are already in the high-temperature regime, but the
temperature is still not too high, such that in the thermal mass
$m_y(T)$ the temperature corrections are subdominant.  In this case,
neglecting the thermal corrections to the mass, we get the estimate for $C_b$,

\begin{equation}
C_b \approx  5.5 \times 10^4 \frac{1}{\lambda_y^4} \frac{m_y^6}{T^6}
{\rm ln}^2\left( 1.2465 m_y/T\right) \;,
\label{Cb}
\end{equation}
recalling that for the model  (\ref{superW}), $\lambda_y = 6
h^2$. Taking $m_y/T \sim 0.1$, we obtain $ 70 \lesssim C_b \lesssim
1.8 \times 10^{4}$, for values of $h$ between $0.1$ and $0.2$. This is
in the absence of further decay modes, which would increase even more
the estimates for $C_b$ (the bulk viscosity coefficient  is
proportional to the radiation bath field degeneracy). These values are
already within the window of values of the  viscosity coefficient
observed by the results in {}Figs. \ref{figch}, \ref{fighy} and
\ref{fighi},  which allows warm inflation with smaller dissipation as
a consequence   of including a bulk viscous pressure.

%%%%%%%%%%%%%%%%%%%%%%%%%%%%%%%%%%%%%%%%%%%%%%%%%%%%

\section{Conclusions}
\label{sec7}

In this work we have examined the stability of the dynamics of warm
inflation when bulk viscous effects are included. The noncausal theory
for the bulk viscous pressure, given by the Eckart hydrodynamics
theory, was studied along with two other causal theories, the IS
linear theory and the NLCDH theory, which is a simpler causal theory
for the bulk viscosity, yet nonlinear. The stability of the resulting
dynamical system for warm inflation has been studied before by the
authors of Refs. \cite{MX}, in the absence of bulk viscous pressure
and  \cite{pavon},  by including the effect of a viscous radiation
bath for only the Eckart case.  In this work we have extended these
previous works by also carrying out the analysis of stability for the
causal theories.  We have also explored model building in the context
of warm inflation with bulk viscous pressure.

By including the analysis of stability for the causal theories of IS
and NLCDH, it becomes possible to clarify on the differences between
these two theories and with that of the noncausal Eckart theory.  We
have seen significant differences for the radiation production in each
of these different theories as the relaxation time of the radiation
fluid increases. Among the three theories for the bulk pressure we
have studied, the NLCDH case proved to be the most robust of them as
far as stability is concerned.  We have shown that it allows for
relatively larger bulk viscous pressures  as compared to the Eckart
and IS cases.  The noncausal theory is the one that most quickly can
develop instabilities. Our results also suggest the significance and
differences there might be in other nonlinear generalizations of
theories for the viscous hydrodynamics of radiation fluids in
cosmology, when relatively larger relaxation times are involved, but
still within the range of validity of thermal equilibrium for the
radiation bath, $\Theta < 1$.   Recalling that the Eckart and the IS
theories have a long history of use in cosmology in
general~\cite{earlierbulk,darkenergy},  other nonlinear realizations
of viscous hydrodynamics, in particular for the bulk pressure, may
lead to significant differences in results, similar to the comparisons
we found here for our problem.

In regards model building, we have shown that accounting for bulk
viscous pressure effects in the radiation fluid can relax the
requirements on the magnitude of the dissipation coefficient for the
inflaton field, especially for a large bulk viscosity coefficient.
This range of bulk viscosity coefficients can be realized within the
regime of stability requirements in warm inflation, and this range is
within reach of realistic model parameters.  The upshot is, accounting
for bulk viscosity effects makes warm inflation more robust.  In
particular warm inflation is more readily realized and with less
constraints when bulk viscosity effects are accounted for versus when
they are ignored.  Although our examination of warm inflation
was only for one particular form of the dissipative coefficient,
we believe this conclusion will hold in general.
Our results point to the importance of considering
viscous effects in future studies of warm inflation.  We also believe
our results can have more general implications for systems involving
viscous radiation fluids, in particular for cosmology.  
Very little has been discussed in the literature on the
significance to cosmology of actual expressions of bulk viscosity computed
from underlying particle physics models.  This is one
of the few papers that has examined this question for the specific
case of warm inflation dynamics. However the methodology developed
here can easily be extended to examine the same sorts of 
bulk viscosity effects
in other relevant cosmological epochs, such as
during the reheating
phase after inflation in cold inflation dynamics and
during the radiation and matter dominated expansion regimes.

\acknowledgements

M.B-G. and R.C. are partially supported by MICINN (FIS2010-17395) and
``Junta de Andaluc\'ia" (FQM101) A.B. acknowledges support from the
STFC. R.O.R is partially supported by research grants from Conselho
Nacional de Desenvolvimento Cient\'{\i}fico e Tecnol\'ogico (CNPq) and
Funda\c{c}\~ao Carlos Chagas Filho de Amparo \`a Pesquisa do Estado do
Rio de Janeiro (FAPERJ). G.S.V. is supported by Coordena\c{c}\~ao de
Aperfei\c{c}oamento de Pessoal de N\'{\i}vel Superior (CAPES).
R.O.R. and G.S.V. would also like to thank T. Koide for discussions concerning
the hydrodynamical equations with bulk viscosity.

%%%%%%%%%%%%%%%%%%%%%%%%%%%%%%%%%%%%%%%%%%%%%%%%%%%%

\end{document}